\documentclass[12pt]{article}
\usepackage{amsfonts}
\usepackage{bm}
\usepackage[dvips]{epsfig}
\usepackage{graphicx}
\usepackage{amssymb}
\usepackage{amsmath}

\begin{document}

\renewcommand{\theequation}{\thesection.\arabic{equation}}

\title{The conservation of the Hamiltonian structures in the
deformations of the Whitham systems. }

\author{A.Ya. Maltsev.}

\date{ \centerline{L.D. Landau Institute for Theoretical Physics}
\centerline{142432 pr. Ak. Semenova 1A, Chernogolovka,
Moscow reg.} 
\centerline{maltsev@itp.ac.ru}}

\maketitle

\begin{abstract}
We consider the construction of the deformed Whitham system for
the KdV-equation in the one-phase case and investigate the
conservation of the Hamiltonian properties in this situation.
It is shown then, that both the Gardner - Zakharov - Faddeev
and the Magri brackets give the deformed Dubrovin - Novikov
brackets (the brackets of Dubrovin - Zhang type) for the deformed
Whitham system constructed by our procedure. The general approach
used in the paper gives a scheme for the averaging of the
Poisson structures in the general situation.
\end{abstract}

\section{Introduction.}

 We are going to consider the conservation of local 
field-theoretical Hamiltonian structures in the method of 
deformations of the Whitham systems. As it is well known,
the Whitham method is connected with the slow modulations of
parameters of (one-phase or multi-phase) periodic or quasi-periodic
solutions of PDE's while the Whitham system itself rules the
behavior of the modulated parameters as of the functions of
time and spatial variables. The Whitham system is usually
written as a system of Hydrodynamic type 

\begin{equation}
\label{WhithamSystem}
U^{\nu}_{T} \,\, = \,\, V^{\nu}_{\mu} ({\bf U}) \, U^{\mu}_{X} 
\end{equation}
and gives the main term in the connection of the time and
spatial derivatives of parameters $U^{\nu} (X, T)$. The variables
$T$ and $X$ represent usually the "slow" time and spatial variables
$T = \epsilon \, t$, $X = \epsilon \, x$ connected with the
variables $t$ and $x$ through the small parameter $\epsilon$.
The Whitham system (\ref{WhithamSystem}) is then a homogeneous 
system of Hydrodynamic Type connecting the first derivatives of
the slow modulated parameters. Different aspects and numerous
applications of the Whitham method were studied in many different
works and the Whitham method is considered now as one of the
classical methods of investigation of non-linear systems.

 Different properties of the Whitham equations were investigated
by many authors (see f.i. \cite{AblBenny} - \cite{AvNov2},
\cite{dm} - \cite{dn3}, \cite{ElKrVen}, \cite{ffm} - \cite{gravtian},
\cite{GurKrEl1} - \cite{LaxLevVen}, \cite{luke} - \cite{malnloc2},
\cite{sympform} - \cite{DefLin}, \cite{Nov} - \cite{tian2}, 
\cite{Ven1} - \cite{theorsol}). Thus, it pointed out 
by G. Whitham (\cite{whith1,whith2,whith3}) that the Whitham system 
(\ref{WhithamSystem}) has a local Lagrangian structure in 
the case when the initial system has a local Lagrangian
structure
 
$$\delta \,\, \int \int
{\cal L} (\bm{\varphi}, \, \bm{\varphi}_{t}, \, \bm{\varphi}_{x} ,
\, \dots ) \,\, dx \, dt \,\,\, = \,\,\, 0 $$
on the initial phase space the space $\{\bm{\varphi}(x,t)\}$.

 The procedure of construction of the Lagrangian formalism for 
the Whitham system (\ref{WhithamSystem}) is given by the averaging 
of the Lagrangian function ${\cal L}$ on the family of $m$-phase
solutions of the initial system. Let us note also that in 
the case of presence of additional parameters $n^{l}$ the additional 
method of the Whitham pseudo-phases should be used.
 
 The important procedure of the averaging of local field-theoretical  
Hamiltonian structures was suggested by B.A. Dubrovin and
S.P. Novikov (\cite{dn1,Nov,dn2,dn3}). The Dubrovin-Novikov procedure
gives the local field-theoretical Hamiltonian formalism for the
Whitham system (\ref{WhithamSystem}) in the case when the initial 
system has a local Hamiltonian formalism of general type.
The Dubrovin-Novikov bracket for the Whitham system has
a general form

\begin{equation}
\label{DubrNovbracket}
\{U^{\nu}(X), \, U^{\mu}(Y) \} \,\,\, = \,\,\,
g^{\nu\mu}({\bf U}) \,\, \delta^{\prime}(X-Y) \,\, + \,\,
b^{\nu\mu}_{\lambda} ({\bf U}) \, U^{\lambda}_{X} \,\,
\delta (X-Y)
\end{equation}
and was called the local Poisson bracket of Hydrodynamic type.
The theory of the brackets (\ref{DubrNovbracket}) is closely related
with differential geometry (\cite{dn1,dn2,dn3}) and is connected
with different coordinate systems in the (pseudo) Euclidean
spaces. Let us say also that during the last years the
important weakly-nonlocal generalizations of Dubrovin-Novikov
brackets (Mokhov-Ferapontov bracket and Ferapontov brackets) 
were introduced and studied
(\cite{mohfer1,fer1,fer2,fer3,fer4,pavlov2,pavlov,PhysD}).

 During the last years the theory of deformations of systems
(\ref{WhithamSystem}) and the Poisson brackets (\ref{DubrNovbracket})
was intensively studied 
(\cite{Dubrov1,Dubrov2,Dubrov3,Dubrov4,DubrZhang1,Dubrov5,   
DubrZhang2,Lorenzoni,DubrZhang3,LiuZhang1,LiuZhang2,DubrLiuZhang,
DubrZhangZuo}). The $\epsilon$-deformations of systems
of Hydrodynamic Type (\ref{WhithamSystem}) and of brackets
(\ref{DubrNovbracket}) give the "dispersive" corrections to
(\ref{WhithamSystem}) and (\ref{DubrNovbracket}) and are
represented usually as the formal series in the powers of
$\epsilon$ with the higher derivatives of the parameters
${\bf U}$. Let us say that the theory of the compatible Poisson
brackets (\ref{DubrNovbracket}) and their deformations
demonstrate very nontrivial structures and is considered
now as one of the general approaches in the classification
of integrable hierarchies.

 We will consider here the deformations of systems
(\ref{WhithamSystem}) and the Poisson brackets (\ref{DubrNovbracket})
connected immediately with the Whitham method for the
slow-modulated parameters.  As far as we know, the idea of 
consideration of dispersive Whitham systems appeared first 
in the paper of M.J. Ablowitz and D.J. Benney (\cite{AblBenny}) 
where the dispersive character of the higher corrections in 
Whitham approach was pointed out. The regular procedure of 
deformation of the Whitham systems was
constructed in \cite{deform1} in connection with the theory
of deformations of systems of Hydrodynamic Type developed in 
\cite{DubrZhang1,DubrZhang2}. In \cite{DefLin} a special 
modification of the deformation procedure which gives a regular 
transition from the linear to non-linear systems was also suggested. 
The procedure used in \cite{DefLin} represents the method of
A.C. Newell giving the connection between the
Whitham approach and the Nonlinear Shr\"odinger equation
approach to the slow modulations (\cite{Newell}) which was modified
to the case of the deformed Whitham systems.

 We will consider in this paper
the modification of the deformation procedure considered in 
(\cite{DefLin}) since the regular properties of the deformation
procedure in the case of the vanishing amplitude of
oscillations seem to be important in many situations.

 The main goal of this paper is to prove the conservation
of the local field-theoretical Hamiltonian structures in the
method of the deformations of the Whitham systems which is
considered in the example of the one-phase modulated solutions
of the KdV equation. Namely, we suggest here a scheme of the
"averaging" of local field-theoretical Poisson brackets giving
the deformed Dubrovin - Novikov brackets for the deformed Whitham
systems (\ref{WhithamSystem}). The procedure considered here is
based on the Dirac procedure of the restriction of the
Poisson bracket on a sub-manifold which provides the Jacobi
identity for the "averaged" Poisson structures.

 In Chapters 2 and 3 we describe the scheme of deformation
of the Whitham system for the KdV equation giving the dispersive
corrections to the standard system of Whitham in this situation.
In Chapter 4 we consider immediately the averaging of two local
Hamiltonian structures for KdV and prove the existence of two
deformed brackets (\ref{DubrNovbracket}) for the deformed
Whitham system. Finally, in Chapter 5, a scheme of the averaging
of the local Lagrangian structures in the method of deformations
of the Whitham system is also considered. The construction used
here have in fact a general character and can be used in analogous
form for different systems of PDE's.

\section{The Whitham method and the deformation scheme.}

 As is well known the Whitham method (\cite{whith1,whith2,whith3})
is connected with the slow modulations of periodic or
quasiperiodic $m$-phase solutions of nonlinear systems

\begin{equation}
\label{insyst}
F^{i}(\bm{\varphi}, \bm{\varphi}_{t}, \bm{\varphi}_{x}, \dots )
\,\, = \,\, 0
\,\,\,\,\,\,\,\, , \,\,\,\,\, i = 1, \dots, n \,\,\, , \,\,\,
\bm{\varphi} = (\varphi^{1}, \dots, \varphi^{n})
\end{equation}
which are represented usually in the form

\begin{equation}
\label{phasesol}
\varphi^{i} (x,t) \,\, = \,\, \Phi^{i} \left( {\bf k}({\bf U})\, x
\, + \, \bm{\omega}({\bf U})\, t \, + \, \bm{\theta}_{0}, \,
{\bf U} \right)
\end{equation}

 In these notations the functions ${\bf k}({\bf U})$ and
$\bm{\omega}({\bf U})$ play the role of the "wave numbers"
and "frequencies" of $m$-phase solutions and $\bm{\theta}_{0}$
are the initial phase shifts. The parameters of the solutions
${\bf U} = (U^{1}, \dots, U^{N})$ can be chosen in arbitrary
way, however, we assume that they do not change under arbitrary
shifts of the initial phases $\bm{\theta}_{0}$ of solutions.

 The functions $\Phi^{i}(\bm{\theta})$ satisfy the system

\begin{equation}
\label{PhaseSyst0}
F^{i} \left( {\bf \Phi}, \omega^{\alpha}
{\bf \Phi}_{\theta^{\alpha}},
k^{\beta} {\bf \Phi}_{\theta^{\beta}},   
\dots \right) \,\,\, \equiv
\,\,\, 0
\,\,\,\,\,\,\,\, , \,\,\,\,\, i = 1, \dots, n
\end{equation}
and we choose for every ${\bf U}$ some function
${\bf \Phi}(\bm{\theta}, {\bf U})$ as having
"zero initial phase shifts". The full set of $m$-phase 
solutions of (\ref{insyst}) can then be represented in form
(\ref{phasesol}). For $m$-phase solutions of (\ref{insyst})
we have then
${\bf k}({\bf U}) = (k^{1}({\bf U}), \dots, k^{m}({\bf U}))$, 
$\bm{\omega}({\bf U}) = (\omega^{1}({\bf U}), \dots,
\omega^{m}({\bf U}))$,
$\bm{\theta}_{0} = (\theta^{1}, \dots, \theta^{m})$,
where ${\bf U}= (U^{1}, \dots, U^{N})$ are parameters
of the solution. We require also that all the functions
$\Phi^{i}(\bm{\theta}, {\bf U})$ are $2\pi$-periodic with
respect to every $\theta^{\alpha}$, $\alpha = 1, \dots, m$.

 Let us denote by $\Lambda$ the family of the functions
$\bm{\Phi}(\bm{\theta}, {\bf U})$ which depend on the parameters
${\bf U}$ in a smooth way and satisfy system (\ref{PhaseSyst0})
for all ${\bf U}$. We will assume also that $\Lambda$ is the
maximal family having these properties.

 In Whitham approach the parameters
${\bf U}$ become slow functions of $x$ and $t$:
${\bf U} = {\bf U}(X,T)$, where $X = \epsilon x$, $T = \epsilon t$
($\epsilon \rightarrow 0$).

 The functions ${\bf U}(X,T)$ should satisfy in this case
some system of differential equations (Whitham system) which
makes possible the construction of the corresponding
asymptotic solution. More precisely (see \cite{luke}), we try  
to find the asymptotic solutions

\begin{equation}
\label{whithsol}
\varphi^{i}(\bm{\theta}, X, T) \,\,\, = \,\,\, \sum_{k\geq0}  
\Psi^{i}_{(k)} \left( {{\bf S}(X,T) \over \epsilon} +
\bm{\theta}, \, X, \, T \right) \,\, \epsilon^{k}
\end{equation}
(where all $\bm{\Psi}_{(k)}$ are $2\pi$-periodic in $\bm{\theta}$)
which satisfy the system (\ref{insyst}), i.e.
 
$$F^{i} \left( \bm{\varphi}, \epsilon \bm{\varphi}_{T},
\epsilon \bm{\varphi}_{X}, \dots \right) \,\,\, = \,\,\, 0
\,\,\,\,\,\,\,\, , \,\,\,\,\, i = 1, \dots, n $$
 
 The function ${\bf S}(X,T) = (S^{1}(X,T), \dots, S^{m}(X,T))$
is called a "modulated phase" of solution (\ref{whithsol}).

 It is easy to see that the function
$\bm{\Psi}_{(0)}(\bm{\theta}, X, T)$ should belong
to the family of $m$-phase solutions of (\ref{insyst})
at every $X$ and $T$. We have then

\begin{equation}
\label{psi0}
\bm{\Psi}_{(0)} (\bm{\theta},X,T) \,\,\, = \,\,\,
\bm{\Phi} \left( \bm{\theta} + \bm{\theta}_{0}(X,T), {\bf U}(X,T)
\right)
\end{equation}
and

$$S^{\alpha}_{T}(X,T) \, = \, \omega^{\alpha}({\bf U}) \,\,\, ,   
\,\,\,\,\, S^{\alpha}_{X}(X,T) \, = \, k^{\alpha}({\bf U}) $$
as follows from the substitution of (\ref{whithsol}) into system
(\ref{insyst}).

 The functions $\bm{\Psi}_{(k)} (\bm{\theta},X,T)$ are defined
from the linear systems
 
\begin{equation}
\label{ksyst}
{\hat L}^{i}_{j[{\bf U}, \bm{\theta}_{0}]}(X,T) \,\,
\Psi_{(k)}^{j} (\bm{\theta},X,T) \,\,\, = \,\,\,  
f_{(k)}^{i} (\bm{\theta},X,T)
\end{equation}
where ${\hat L}^{i}_{j[{\bf U}, \bm{\theta}_{0}]}(X,T)$
is a linear operator given by the linearization of system
(\ref{PhaseSyst0}) on solution (\ref{psi0}). The resolvability
conditions of systems (\ref{ksyst}) can be written as the
orthogonality conditions of the functions
${\bf f}_{(k)} (\bm{\theta},X,T)$ to all the "left eigen vectors" 
(the eigen vectors of adjoint operator) of the operator
${\hat L}^{i}_{j[{\bf U}, \bm{\theta}_{0}]}(X,T)$ corresponding to
zero eigen-values. The resolvability conditions of (\ref{ksyst})
for $k = 1$

\begin{equation}
\label{1syst}
{\hat L}^{i}_{j[{\bf U}, \bm{\theta}_{0}]}(X,T) \,\,
\Psi_{(1)}^{j} (\bm{\theta},X,T) \,\,\, = \,\,\,
f_{(1)}^{i} (\bm{\theta},X,T)
\end{equation}
together with the relations
$k^{\alpha}_{T} = \omega^{\alpha}_{X} $
give the Whitham system for $m$-phase solutions of (\ref{insyst})
which plays the central role in the slow modulations approach.

 Let us say that the resolvability conditions of (\ref{ksyst})
can in fact be rather complicated in a general multi-phase   
case. Indeed, we need to investigate the eigen-spaces
of the operators ${\hat L}_{[{\bf U}, \bm{\theta}_{0}]}$
and ${\hat L}^{\dagger}_{[{\bf U}, \bm{\theta}_{0}]}$
on the space of $2\pi$-periodic functions which can be rather
non-trivial in the multi-phase situation. Thus even the dimensions
of kernels of ${\hat L}_{[{\bf U}, \bm{\theta}_{0}]}$
and ${\hat L}^{\dagger}_{[{\bf U}, \bm{\theta}_{0}]}$
can depend in non-smooth way on the values of ${\bf U}$
so we can have a rather complicated picture on
the ${\bf U}$-space (\cite{dm,dobr1,dobr2}).

 These difficulties do not usually appear in the one-phase
situation ($m = 1$) where the behavior of eigen-values of
${\hat L}_{[{\bf U}, \bm{\theta}_{0}]}$
and ${\hat L}^{\dagger}_{[{\bf U}, \bm{\theta}_{0}]}$
is usually rather regular.

 In this chapter we are going to consider a scheme of deformation
of the Whitham system giving "dispersive" corrections
to the system of Hydrodynamic Type which describe in fact the
higher corrections to the corresponding asymptotic solutions.
We are going to use here the one-phase modulated solutions of
the KdV equation as a basic example throughout the paper, so
let us consider now the KdV equation

\begin{equation}
\label{kdv}
\varphi_{t} \, + \, \varphi \varphi_{x} \, + \,
\varphi_{xxx} \,\, = \,\, 0
\end{equation}

It has a family of exact solutions of the form
\begin{equation}
\label{phiphi}
\varphi(x,t) = \Phi(kx + \omega t; U^{1}, U^{2}, U^{3})
\end{equation}
where the functions $\Phi(\theta; U^{1}, U^{2}, U^{3})$  
depending on three real parameters $U^{1}, U^{2}, U^{3}$ 
is $2\pi$-periodic in $\theta$, the wave 
number $k$ and the frequency $\omega$ are uniquely determined 
by these parameters,
$$ k \, = \, k (U^{1}, U^{2}, U^{3}) \,\,\,\,\, , \,\,\,\,\,
\omega \, = \, \omega (U^{1}, U^{2}, U^{3}) $$
 As we pointed out already the Whitham modulation theory gives 
a prescription for finding approximate solutions to KdV in the form
\begin{equation}
\label{whi1}
\varphi \simeq \Phi \left(
{S(X,T) \over \epsilon}; U^{1}(X,T), U^{2}(X,T), 
U^{3}(X,T) \right)
\end{equation}
where $\epsilon$ is a small parameter, 
$$ X \, = \, \epsilon\, x \,\,\, , \,\,\, T \, = \, \epsilon\, t $$
are slow variables, the dependence of the parameters 
$U^{\nu} = U^{\nu} (X,T)$ is 
determined from certain hyperbolic system of the first order 
quasilinear equations of the form
\begin{equation}
\label{whi2}
U^{\nu}_{T} \,\, = \,\,  V^{\nu}_{\mu}(U^{1}, U^{2}, U^{3}) \,\,
U^{\mu}_{X} \,\,\,\,\, , \,\,\,\,\, 
\nu \, , \, \mu \, = \, 1, \dots , 3
\end{equation}

The phase function $S(X,T)$ is determined by quadratures
\begin{eqnarray}
\label{whi3}
&&
\partial_{X} S(X,T) = k (U^{1}(X,T), U^{2}(X,T), U^{3}(X,T))
\nonumber\\
&&
\\
&& \partial_{T} S(X,T) = \omega (U^{1}(X,T), U^{2}(X,T), U^{3}(X,T))
\nonumber
\end{eqnarray}
We now want to describe the higher corrections to the approximate 
solutions \eqref{whi1}. We will call them {\it deformed Whitham 
equations}.

The deformed Whitham equations will arise in the description 
of solutions to \eqref{kdv} in the form
\begin{equation}
\label{kdvexp}
\varphi = \Phi\left({S(X,T)}+\theta; 
{\bf U}(X,T)\right) \, + \, \sum_{l\geq 1} 
\Phi_{(l)}(S(X,T) +\theta; X, T)
\end{equation}
where the functions 
$$\Phi_{(l)}(\theta; X, T) \, = \, 
\Phi_l (\theta; {\bf U}, {\bf U}_X, {\bf U}_{XX}, 
\dots, {\bf U}^{(l)}) $$ 
$2\pi$-periodic in $\theta$ are graded homogeneous differential 
polynomials in ${\bf U}_X$, ${\bf U}_{XX}$ etc. with coefficients 
being smooth functions of ${\bf U}=(U^{1}, U^{2}, U^{3})$. 
The gradation is defined by the rule
$$
\deg \partial_{X}^{m} {\bf U} \, = \, m, \quad m = 1, 2, \dots.
$$
As usual the degree of the product of homogeneous differential 
polynomials is equal to the sum of their degrees. We use here the 
notations $X$ and $T$ just to emphasize that the 
functions ${\bf U}(X,T)$ are "slow" functions of spatial and
time variables. At the moment we do not write the small parameter 
$\epsilon$ explicitly; it will be reintroduced later.

\begin{figure}
\begin{center}
\epsfig{file=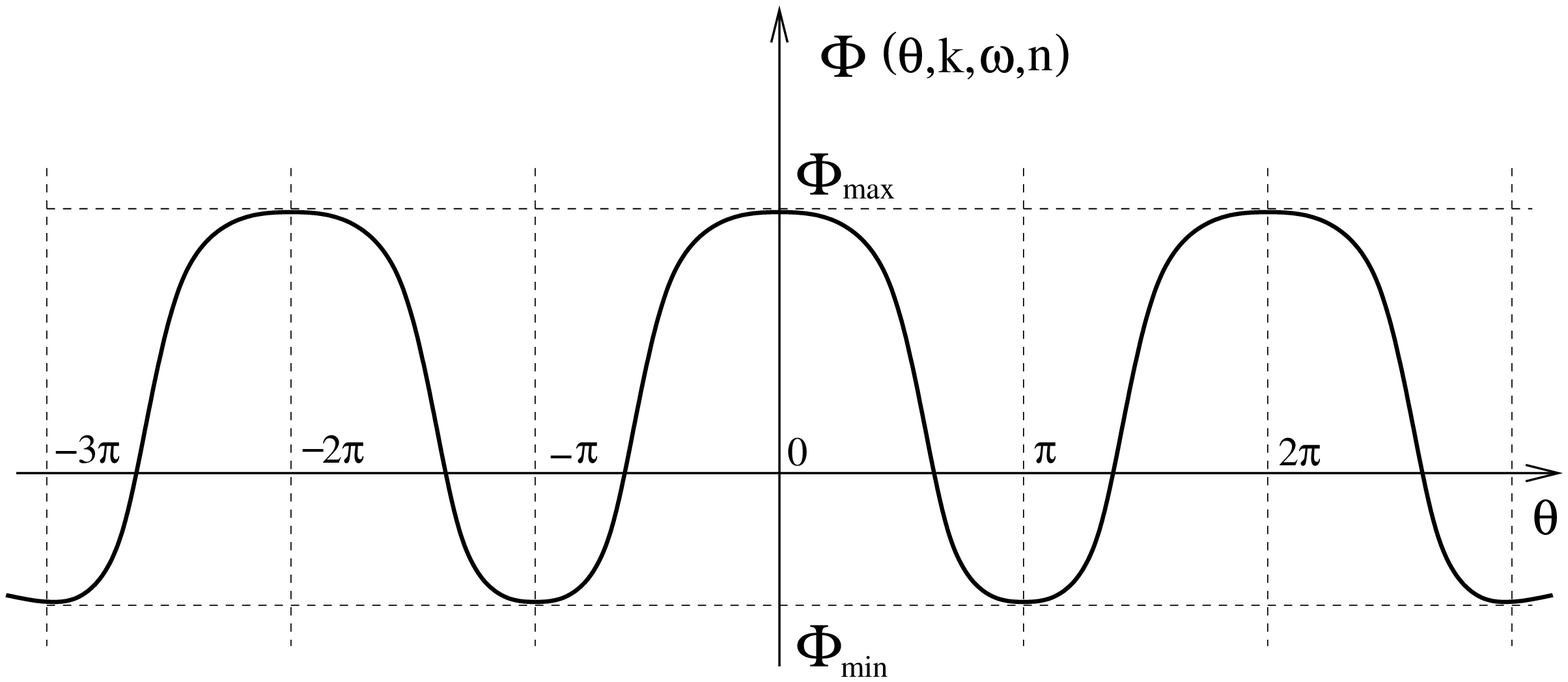,width=14.0cm,height=7cm}
\end{center}
\caption{The function $\Phi(\theta,k,\omega)$ having zero initial
phase shift.}
\label{zerophase}
\end{figure}

It will be convenient to choose a particular system of 
coordinates $U^{1}, U^{2}, U^{3}$ in the space 
of traveling wave solutions $\Phi(\theta; {\bf U})$. 
We denote them
$$
{\bf U}=(k, \omega, n)
$$
where $k$ and $\omega$ are the wave number and the frequency and 
$n$ is the mean value of $\Phi$. The ODE for the function $\Phi$ 
\begin{equation}
\label{phasesyst}
\omega \Phi_{\theta} + k \Phi \Phi_{\theta} +
k^{3} \Phi_{\theta\theta\theta} 
\,\, = \,\, 0
\end{equation}
can be integrated by quadratures
$$
\sqrt{{k^{3} \over 2}} \int_{a_{3}} 
{d \Phi \over 
\sqrt{- k \Phi^{3}/6 - \omega \Phi^{2}/2 + A \Phi + B}}
q\,\,\, = \,\,\, \theta $$
where $a_{3}$ is the third zero of the cubic polynomial
$- k \Phi^{3}/6 - \omega \Phi^{2}/2 + A \Phi + B$ according to
the normalization shown at Fig. \ref{zerophase}. The dependence on the 
parameters of the coefficients of the polynomial $A=A(k,\omega,n)$ 
$B=B(k,\omega,n)$ is determined from the equations
$$\sqrt{{k^{3} \over 2}} \oint
{d \Phi \over 
\sqrt{ - k \Phi^{3}/6 - \omega \Phi^{2}/2 + A \Phi + B}}
\,\,\, = \,\,\, 2\pi $$
$$\sqrt{{k^{3} \over 2}} \oint
{\Phi\, d \Phi \over 
\sqrt{ - k \Phi^{3}/6 - \omega \Phi^{2}/2 + A \Phi + B}}
\,\,\, = \,\,\, 2\pi\, n. $$
We also fix the initial phase shift of the functions 
$\Phi(\theta,k,\omega,n)$
in such a way that every $\Phi(\theta,k,\omega,n)$ has a local maximum
at the point $\theta = 0$ (see Fig. \ref{zerophase}).

 It is well-known that the function 
$- \Phi (k x + \omega t, {\bf U}) / 6 $ represents the one-gap
potential for the Shr\"odinger operator

\begin{equation}
\label{ShrOp}
{\hat L} \,\, = \,\, - \, {d \over d x^{2}} \, - \,
{\varphi \over 6}
\end{equation}
while the KdV equation can be written in the Lax representation

$${d {\hat L} \over d t} \,\, = \,\, \left[
{\hat A} , {\hat L} \right] $$
where

$${\hat A} \,\, = \,\, - \, 4 \, {d^{3} \over d x^{3}}
\, - \, {1 \over 2} \, \left( \varphi \, {d \over d x} \, + \,
{d \over d x} \, \varphi \right) $$

 Let us say that the integrability of the KdV equation will be
convenient in some aspects of our considerations. However, the
general questions considered here are not connected with the
integrability and are applicable for the non-integrable examples
as well. 

 It is well-known also that the solution
$\Phi (k x + \omega t, {\bf U})$ can be represented in the form

$$\Phi (k x + \omega t, {\bf U}) \,\, = \,\, {2 a \over s^{2}}
\, {\rm dn}^{2} \left[ \left( {a \over 6 s^{2}}\right)^{1/2}
(x - V t), s \right] \, + \, \gamma $$
$$V \,\, = \,\, {2 a \over 3 s^{2}} (2 - s^{2}) \, + \,
\gamma $$
where $s$ is the modulus of the Jacobi elliptic function
${\rm dn} (u, s)$, $0 \leq s \leq 1$. The value $2 a$ plays
the role of the amplitude of oscillations for the periodic 
solution and the values $(k, \omega, n)$ can be expressed in 
terms of the parameters $(a, s, \gamma)$ in the following
way

$$ k \, = \, {\pi \over K (s)} \left( 
{a \over 6 s^{2}}\right)^{1/2} \,\,\, , \,\,\,
\omega \, = \, - \, V \, k \, = \, - \,
{4 \pi \over K (s)} (2 - s^{2})
\left( {a \over 6 s^{2}}\right)^{3/2} \, - \,
{\gamma \pi \over K (s)} \left( {a \over 6 s^{2}}\right)^{1/2} $$
$$n \, = \, \gamma \, + \, {2 a E (s) \over s^{2} K (s)} $$
where $K (s)$ and $E (s)$ are the elliptic integrals of the first
and the second kind respectively.

 We can write also

$$\Phi (\theta, a, s, \gamma) \,\, = \,\,
{2 a \over s^{2}} \, {\rm dn}^{2} \left(
{K (s) \over \pi} \theta, s \right) \, + \, \gamma $$
for our normalization of the functions $\Phi (\theta, {\bf U})$.
Let us note also that the parameters $(a, s, \gamma)$ are connected
with the energy band edges $(r_{1}, r_{2}, r_{3})$ of the operator
(\ref{ShrOp}) by the formulas

$$r_{2} \, - \, r_{1} \,\, = \,\, a \,\,\, , \,\,\,
{r_{2} - r_{1} \over r_{3} - r_{1}} \,\, = \,\, s^{2}
\,\,\, , \,\,\, r_{1} + r_{2} - r_{3} \,\, = \,\, \gamma $$
($r_{3} > r_{2} > r_{1}$).

The total function $\Phi^{(tot)}(\theta,X,T)$:

$$\Phi^{(tot)}(\theta,X,T) \,\,\, = \,\,\, 
\sum_{l \geq 0} \Phi_{(l)} (\theta,X,T) \,\,\, = \,\,\,
\phi(\theta - S(X,T), X, T) $$
satisfies the equation

$$S_{T} \Phi^{(tot)}_{\theta} \, + \, S_{X} 
\Phi^{(tot)} \Phi^{(tot)}_{\theta} \, + \,
\left( S_{X} \right)^{3} \Phi^{(tot)}_{\theta\theta\theta} \, + $$
$$\, + \Phi^{(tot)}_{T} \, + \, 
\Phi^{(tot)} \Phi^{(tot)}_{X} \, + \, 
3 S_{X}^{2} \Phi^{(tot)}_{\theta\theta X} \, + \, 
3 S_{X} S_{XX} \Phi^{(tot)}_{\theta\theta} \, + $$
\begin{equation}
\label{fullsyst}
+ \, 3 S_{X} \Phi^{(tot)}_{\theta XX} \, + \, 
3 S_{XX} \Phi^{(tot)}_{\theta X}
\, + \, S_{XXX} \Phi^{(tot)}_{\theta} \, + 
\end{equation}
$$+ \, \Phi^{(tot)}_{XXX} \,\,\, = \,\,\, 0 $$
This  yields linear equations for the functions $\Phi_l(\theta; {\bf u})$ 
for $l\geq 1$. In particular the function $\Phi_{(1)}(\theta,X,T)$ 
satisfies the equation
\begin{equation}
\label{firstappr}
\omega \Phi_{(1)\theta} \, + \, k \Phi_{(1)} \Phi_{\theta} 
\, + \,  k \Phi_{(1)\theta} \Phi \, + \, k^{3}
\Phi_{(1)\theta\theta\theta} \,\, = \,\, f_{(1)} (\theta, X, T)
\end{equation}
where
\begin{equation}
\label{f1}
f_{(1)} (\theta, X, T) \,\, = \,\, - \, \Phi_{T}^{[1]} \, - \,
\Phi \Phi_{X} \, - \, 3 k^{2} \Phi_{\theta\theta X} \, - \,
3 k k_{X} \Phi_{\theta\theta}
\end{equation}
and the notation $^{[1]}$ means that we consider just the part
of $\Phi_{T}$ having degree $1$ according to our definition.

Denote ${\hat L}_{[X,T]}$ the linear operator
\begin{equation}
\label{Loper}
{\hat L}_{[X,T]} \,\, = \,\, \omega \, 
{\partial \over \partial \theta}
\, + \, k \, {\partial \over \partial \theta} \, \Phi \, + \, 
k^{3} \, {\partial^{3} \over \partial \theta^{3}}
\end{equation}
We can rewrite (\ref{f1}) in the form
$${\hat L}_{[X,T]} \, \Phi_{(1)} \,\, = \,\, f_{(1)} $$
In the same way we have the analogous systems for the
functions $\Phi_{(l)}(\theta,X,T)$ having the form
\begin{equation}
\label{lsyst}
{\hat L}_{[X,T]} \, \Phi_{(l)} \,\, = \,\,
\omega \Phi_{(l)\theta} \, + \, k \Phi_{(l)} \Phi_{\theta}
\, + \, k \Phi_{(l)\theta} \Phi \, + \, k^{3}
\Phi_{(l)\theta\theta\theta} \,\, = \,\, f_{(l)} (\theta, X, T)
\end{equation}
where $f_{(l)} (\theta, X, T)$ are the discrepancies having degree 
$l$.

 The functions $k(X,T) = S_{X}$, $\omega(X,T) = S_{T}$ and
$n(X,T)$ must satisfy the ``deformed Whitham system"

\begin{eqnarray}
\label{konsyst}
&&
k_{T} \,\,\, = \,\,\, \omega_{X}
\nonumber\\
&&
\nonumber\\
&&
\omega_{T} \,\,\, = \,\,\, \sum_{l \geq 1}
\sigma_{(l)} \left( k, \omega, n,
k_{X}, \omega_{X}, n_{X}, \dots \right)
\\
&&
n_{T} \,\,\, = \,\,\, \sum_{l \geq 1}
\eta_{(l)} \left( k, \omega, n,
k_{X}, \omega_{X}, n_{X}, \dots \right)
\nonumber
\end{eqnarray}
where all $\sigma_{(l)}$, $\eta_{(l)}$ are graded homogeneous 
differential polynomials in
$(k, \omega , n, k_{X}, \omega_{X},
n_{X}, \dots )$ of the degree $l$.

It is easy to see that relations (\ref{konsyst}) give in fact a 
possibility to represent in the form of homogeneous differential 
polynomials any expression
like $k_{TX \dots X}$, $\omega_{TX \dots X}$, $n_{TX \dots X}$,
and even $k_{T \dots TX \dots X}$, 
$\omega_{T \dots TX \dots X}$, $n_{T \dots TX \dots X}$
iterating the subsequent substitution of the series (\ref{konsyst}). 
(The last property will not be necessary for the KdV equation).

According to (\ref{konsyst}) all the time derivatives like
$\Phi_{T}$, $\Phi_{(l)T}$ can also be represented as the sum of 
homogeneous components
$$\Phi_{(l)T} \, = \, \Phi_{(l)T}^{[l]} \, + \, 
\Phi_{(l)T}^{[l+1]} \, + \, \Phi_{(l)T}^{[l+2]} \, + \, \dots $$
where the functions $\Phi_{(l)T}^{[s]}$ are differential polynomials 
of $(k, \omega , n, k_{X}, \omega_{X}, n_{X}, \dots )$ 
of the degree $s$.  

We impose the following orthogonality conditions on the discrepancies 
$f_{(l)} (\theta, X, T)$
\begin{equation}
\label{ortcond}
\int_{0}^{2\pi} f_{(l)} \,\, 
{d \theta \over 2\pi} \,\,\, = \,\,\, 0 \,\,\,\,\,\,\,\, ,
\,\,\,\,\,\,\,\, \int_{0}^{2\pi} \Phi \,\, f_{(l)} \,\,
{d \theta \over 2\pi} \,\,\, = \,\,\, 0
\end{equation}
and also the ``normalization" conditions
\begin{equation}
\label{normcond}
\int_{0}^{2\pi} \Phi_{\theta} \,\, \Phi_{(l)} \,\,
{d \theta \over 2\pi} \,\,\, = \,\,\, 0 \,\,\,\,\,\,\,\, ,
\,\,\,\,\,\,\,\, \int_{0}^{2\pi} \Phi_{(l)} \,\,
{d \theta \over 2\pi} \,\,\, = \,\,\, 0
\end{equation}
for the functions $\Phi_{(l)}(\theta, X, T)$ defined 
from (\ref{lsyst}) modulo the linear combinations 
\linebreak $a(X,T) \Phi_{\theta} + b(X,T) \Phi_{n}$.

 For determination of $\sigma_{(l)}$, $\eta_{(l)}$ we use the
system (\ref{konsyst}) to remove all time derivatives of
$(k, \omega, n)$ after the substitution of (\ref{kdvexp}) into
(\ref{kdv}) in order to represent (\ref{fullsyst}) in the
graded form.

 The functions $\sigma_{(l)}$, $\eta_{(l)}$ arising in
(\ref{konsyst}) are found from the compatibility
conditions of systems (\ref{lsyst}) in the $l$-th order.
It can be shown that conditions (\ref{ortcond})-(\ref{normcond})
define uniquely all the expressions $\sigma_{(l)}$, $\eta_{(l)}$
and the corrections $\Phi_{(l)}$, $l \geq 1$.

\vspace{0.5cm}

So, our prescription for deriving the system (\ref{konsyst}) is 
based on the following three conditions:

I) All the functions $\Phi (\theta; k, \omega, n)$ are chosen
in the way shown at Fig. 1 ;

II) The modulated phase $S (X,T)$ is connected with the
parameters $(k, \omega, n)$ by the relations

$$S_{T} (X,T) \, = \, \omega (X,T) \,\,\, , \,\,\,
S_{X} (X,T) \, = \, k (X,T). $$

III) All the higher corrections $\Phi_{(l)} (\theta, X, T)$,
$l \geq 1$ satisfy the normalization conditions (\ref{normcond}).

\vspace{0.5cm}

 According to the statements above 
system (\ref{konsyst}) is uniquely defined by the conditions 
(I)-(III).

Let us now say some words about solutions of the
system (\ref{lsyst}). We have
\begin{equation}
\label{intlsyst}
\omega \Phi_{(l)} \, + \, k \Phi_{(l)} \Phi \, + \, 
k^{3} \Phi_{(l)\theta\theta} \,\, = \,\, \int^{\theta} 
f_{(l)} (\theta^{\prime}, X, T) \, d \theta^{\prime} \, + \,
\xi_{1} 
\end{equation}
Put $\Phi_{(l)} = \alpha_{(l)} \Phi_{\theta}$
to arrive at
$$k^{3} \alpha_{(l)\theta\theta} \Phi_{\theta} \, + \,
2 k^{3} \alpha_{(l)\theta} \Phi_{\theta\theta} \, = \,
\int^{\theta} f_{(l)} \, d \theta^{\prime} \, + \, \xi_{1}. $$
So
$$k^{3}  \alpha_{(l)\theta} \left( \Phi_{\theta} \right)^{2}
\, = \, \int^{\theta} \Phi_{\theta^{\prime}} \left[
\int^{\theta^{\prime}} f_{(l)} \, d \theta^{\prime\prime} 
\, + \, \xi_{1} \right] d \theta^{\prime} \, + \, \xi_{2} $$
and
\begin{eqnarray}
\label{Phil}
&&
\Phi_{(l)} \, = \, {1 \over k^{3}} \, \Phi_{\theta} \,
\left[ \int^{\theta} {1 \over (\Phi_{\theta^{\prime}})^{2}}
\int^{\theta^{\prime}} \Phi_{\theta^{\prime\prime}}
\int^{\theta^{\prime\prime}} f_{(l)} \, 
d \theta^{\prime\prime\prime} \, d \theta^{\prime\prime} \,
d \theta^{\prime}\right.
\nonumber\\
&&
\\
&& \left.
\quad\quad + \, \xi_{1} \int^{\theta} 
{\Phi \over (\Phi_{\theta^{\prime}})^{2}} d \theta^{\prime} 
\, + \, \xi_{2} \int^{\theta} 
{d \theta^{\prime} \over (\Phi_{\theta^{\prime}})^{2}} \, + \,
\xi_{3} \right].
\nonumber 
\end{eqnarray}
Formula (\ref{Phil}) has a local character and
we have to investigate solution (\ref{Phil}) on the whole
axis $- \infty < \theta < + \infty $. (The expression 
$1/(\Phi_{\theta})^{2}$ has singularities at the points 
$\theta_{n} = \pi n$, $n \in \mathbb{Z}$).
Two important cases can be pointed out in our situation:

I) The function $f_{(l)}(\theta)$ is even,
$f_{(l)}(-\theta) = f_{(l)}(\theta)$;

II) The function $f_{(l)}(\theta)$ is odd,
$f_{(l)}(-\theta) = - f_{(l)}(\theta)$.

Proofs of the following two propositions are straightforward.

\vspace{0.5cm}

{\bf Proposition 2.1.}

{\it For an even smooth periodic $f_{(l)}(\theta)$ the
corresponding solution $\Phi_{(l)}(\theta)$ of (\ref{lsyst})
satisfying conditions (\ref{normcond}) is an odd smooth
periodic function. }

\vspace{0.5cm}

{\bf Proposition 2.2.}

{\it For an odd smooth periodic $f_{(l)}(\theta)$ 
the corresponding solution $\Phi_{(l)}(\theta)$ of (\ref{lsyst})
satisfying conditions (\ref{normcond}) is an even smooth
periodic function. }

\vspace{0.5cm}

In particular the function $f_{(1)}(\theta,X,T)$ is given by the 
following expression
$$ - \, \left[ \Phi_{(0)T} \right]^{[1]} \, - \,
\Phi_{(0)} \Phi_{(0)X} \, - \,
3 S_{X}^{2} \Phi_{(0)\theta\theta X} \, - \,  
3 S_{X} S_{XX} \Phi_{(0)\theta\theta}. $$
Recall that the mark $^{[1]}$ means that we collect
the terms of the degree $1$ . 

 Orthogonality conditions (\ref{ortcond}) can be written as 

$$\int_{0}^{2\pi} \Phi_{(0)\omega} {d \theta \over 2\pi} \,\,
\sigma_{(1)} \,\,\, + \,\,\,
\int_{0}^{2\pi} \Phi_{(0)n} {d \theta \over 2\pi} \,\,  
\eta_{(1)} \,\,\, + \,\,\,
\int_{0}^{2\pi} \Phi_{(0)k} {d \theta \over 2\pi} \,\,
\omega_{X} \,\,\, = \,\,\,
\int_{0}^{2\pi} f_{(1)}^{\prime} {d \theta \over 2\pi} $$
and

$$\int_{0}^{2\pi} \Phi_{(0)} 
\Phi_{(0)\omega} {d \theta \over 2\pi} \,\,  
\sigma_{(1)} \,\,\, + \,\,\,
\int_{0}^{2\pi} \Phi_{(0)}
\Phi_{(0)n} {d \theta \over 2\pi} \,\,
\eta_{(1)} \,\,\, + \,\,\,
\int_{0}^{2\pi} \Phi_{(0)}
\Phi_{(0)k} {d \theta \over 2\pi} \,\,
\omega_{X} \,\,\, = \,\,\,
\int_{0}^{2\pi} f_{(1)}^{\prime} {d \theta \over 2\pi} $$
where

$$f_{(1)}^{\prime} \,\,\, = \,\,\,
- \Phi_{(0)} \Phi_{(0)X} \, - \,
3 S_{X}^{2} \Phi_{(0)\theta\theta X} \, - \,
3 S_{X} S_{XX} \Phi_{(0)\theta\theta} $$

 The equations written above determine the functions
$\sigma_{(1)}(k,\omega,n,k_{X},\omega_{X},n_{X})$ and
\linebreak $\eta_{(1)}(k,\omega,n,k_{X},\omega_{X},n_{X})$.  
In this way we arrive at the standard Whitham system 
\eqref{whi2} as the zero order approximation of (\ref{konsyst}).

More generally, according to our approach the functions
$f_{(l)}(\theta,X,T)$ will be always represented in the form
$$f_{(l)}  \,\,\, = \,\,\, - \, 
\left[ \Phi_{(0)T} \right]^{[l]} \,+ \, f_{(l)}^{\prime}
\,\,\, = \,\,\, - \, \Phi_{(0)\omega} \, \sigma_{(l)} 
\,\, + \,\, \Phi_{(0)n} \, \eta_{(l)} \,\, + \,\, 
f_{(l)}^{\prime} $$
where $f_{(l)}^{\prime}$ does not contain the terms
$\sigma_{(l)}$, $\eta_{(l)}$. The corresponding orthogonality
conditions (\ref{ortcond}) recursively determine all the 
terms $\sigma_{(l)}$, $\eta_{(l)}$. 

It is easy to see that the function $f_{(1)}$ is even: 
$f_{(1)}(-\theta) = f_{(1)}(\theta)$.
We obtain therefore that the function $\Phi_{(1)}(\theta)$
is odd
$\Phi_{(1)}(-\theta) = - \Phi_{(1)}(\theta)$.

Furthermore, a direct substitution gives 
$$f_{(2)}  \,\,\, = \,\,\, - \,
\left[ \Phi_{(0)T} \right]^{[2]} \,+ \, f_{(2)}^{\prime}
\,\,\, = \,\,\, - \, \Phi_{(0)\omega} \, \sigma_{(2)}   
\,\, + \,\, \Phi_{(0)n} \, \eta_{(2)} \,\, + \,\,
f_{(2)}^{\prime} $$
where $f_{(2)}^{\prime}(\theta)$ is odd.

Using equations (\ref{ortcond}) for $l = 2$ we get
immediately $\sigma_{(2)} \equiv 0$, 
$\eta_{(2)} \equiv 0$
for the next terms in (\ref{konsyst}). The total function
$f_{(2)}(\theta,X,T)$ becomes then an odd
function in $\theta$. Hence the second correction
$\Phi_{(2)}(\theta)$ is even. By
simple induction we obtain the following Lemma:
 
\vspace{0.5cm}

{\bf Lemma 2.1.}
 
{\it For the choice of the functions
$\Phi(\theta; k, \omega, n)$ corresponding to
Fig. \ref{zerophase} the following statements are true:

1) All the even terms $\sigma_{(2l)}(k, \omega, n, \dots)$,
$\eta_{(2l)}(k, \omega, n, \dots)$
in the deformation of Whitham system (\ref{konsyst}) are
identically zero: $\sigma_{(2l)} \equiv 0$,
$\eta_{(2l)} \equiv 0$;

2) All odd corrections $\Phi_{(2l+1)}(\theta,X,T)$,  
$l \geq 0$ in (\ref{kdvexp}) are odd functions in $\theta$;

3) All even corrections $\Phi_{(2l)}(\theta,X,T)$, 
$l \geq 1$ in (\ref{kdvexp}) are even functions in $\theta$. }

\section{Deformation scheme for the case of small 
amplitude oscillations.}
\setcounter{equation}{0}

The above procedure of deformation has one weak point. Namely, 
in the procedure described the higher corrections 
$\Phi_{(l)}(\theta, X, T)$ as well as the higher deformation
terms in system (\ref{konsyst}) are singular in the
limit of small amplitude oscillations of $\varphi(x,t)$. 
The reason for such a singular behavior can be explained in 
the following way.

Let us rewrite system (\ref{lsyst}) in form 
(\ref{intlsyst}) i.e.
$$ \omega \Phi_{(l)} \, + \, k \Phi_{(l)} \Phi \, + \,
k^{3} \Phi_{(l)\theta\theta} \,\, = \,\, 
g_{(l)} (\theta, X, T) $$
where the right-hand part $g_{(l)}$ given by the expression
$$ g_{(l)} (\theta, X, T) \, \equiv \, \int_{0}^{\theta} 
f_{(l)} (\theta^{\prime}, X, T) \, d \theta^{\prime} \,\, + \,\,
\delta_{(l)} (X,T) $$
is periodic in $\theta$ due to the conditions (\ref{ortcond}).

We can rewrite this system in the form
\begin{equation}
\label{Phigsyst}
{\hat Q}_{[X,T]} \Phi_{(l)} \, = \, g_{(l)}
\end{equation}
where
\begin{equation}
\label{Qoper}
{\hat Q}_{[X,T]} \, \equiv \, \omega (k, A, n) \, + \,
k \, \Phi \, + \, k^{3} \, 
{\partial^{2} \over \partial \theta^{2}} 
\end{equation}
is a self-adjoint operator on the space of $2\pi$-periodic functions.

 Operator (\ref{Qoper}) has just one eigen-vector 
$\Phi_{\theta}$ with zero eigenvalue on the space of 
$2\pi$-periodic functions. The constants 
$\delta_{(l)} (X,T)$ are uniquely determined by the second condition
(\ref{normcond}) for the solutions $\Phi_{(l)}$ of (\ref{Phigsyst}).
It is not difficult to get analytic expressions for
$\delta_{(l)} (X,T)$. It is also easy to see that 
$\delta_{(l)} (X,T) \equiv 0$ for $l = 2s + 1$, $s \geq 0$.

Provided that conditions (\ref{ortcond}) are satisfied we can
write the solution of (\ref{Phigsyst}) in the form
\begin{equation}
\label{Philsol}
\Phi_{(l)} \, = \, \sum_{j} {1 \over \lambda_{j}} \,
\xi_{j} (\theta, X, T) \, \langle \xi_{j} , g_{(l)} \rangle
\end{equation}
where $\xi_{j} (\theta, X, T)$ are the normalized eigen-vectors
of ${\hat Q}_{[X,T]}$ corresponding to non-zero eigenvalues
$\lambda_{j}$.

 Let us consider now operator (\ref{Qoper}) for the case of
small amplitude oscillations:
$$\Phi (\theta, X, T) \, = \, n (X,T) \, + \, a_{0} (X,T) \,
\cos \theta \, + \, \dots \,\,\,\,\,\,\,\, , \,\,\,\,\,
a_{0} \rightarrow 0 $$
The parameter $a_{0} (X,T)$ is the amplitude of the first Fourier
harmonic of $\Phi (\theta, X, T)$ which is similar to the parameter
$A = \Phi_{max} - \Phi_{min}$ in the limit 
$A \rightarrow 0$.

 Operator (\ref{Qoper}) has always the eigen-vector 
$\xi (\theta, X, T) = \Phi_{\theta} (\theta, X, T)$
corresponding to zero eigenvalue, which corresponds to the
function $ - \sin \theta $ in the limit $A \rightarrow 0$.
The dispersion relation $\omega = \omega (k, A, n)$ becomes
the dispersion relation of the linear system
$$\omega \, = \, - \, n \, k \, + \, k^{3} $$
for $A = 0$.

However, the function $\cos \theta$ gives also an eigen-vector
of linear operator $(A =0)$ (\ref{Qoper}) corresponding to zero
eigenvalue. As a result there exists an eigen-vector 
$\xi_{1} (\theta, X, T)$ of the operator ${\hat Q}_{[X,T]}$
corresponding to ``small" eigenvalue 
$\lambda_{1} \rightarrow 0$ (for $A \rightarrow 0$).

 The values $\xi_{1} (\theta )$ and $\lambda_{1}$ can be
also expressed in terms of the elliptic functions in our case.
Indeed, if we compare the operator $- {\hat Q}_{[X,T]} / k $
with the Shr\"odinger operator (\ref{ShrOp}) we can easily
see that the operator $- {\hat Q}_{[X,T]} / k $ is represented
by the Shr\"odinger operator with a one-zone potential multiplied
by $6$. It is well known 
(see f.e. \cite{CoopKharSukh,BelEnol}) that the operator
$- {\hat Q}_{[X,T]} / k $ represents a 3 - energy gaps
Shr\"odinger operator with an elliptic potential in this case.

\begin{figure}
\begin{center}
\epsfig{file=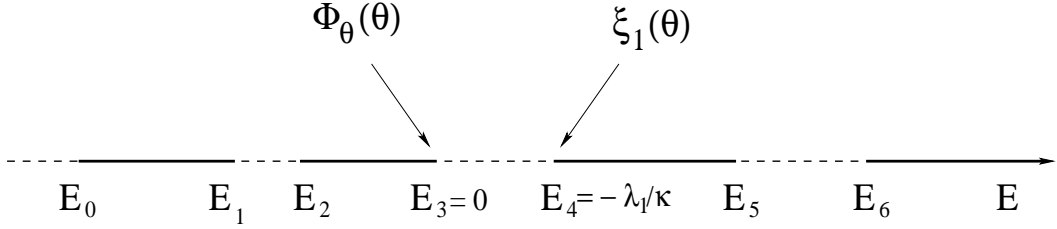,width=14.0cm,height=3cm}
\end{center}
\caption{The spectrum of the operator $- {\hat Q}_{[X,T]} / k $.
The intervals $[E_{0}, E_{1}]$ , $[E_{2}, E_{3}]$, $[E_{4}, E_{5}]$
and $[E_{1}, E_{2}]$, $[E_{3}, E_{4}]$, $[E_{5}, E_{6}]$ represent
the energy bands and the energy gaps of a finite size respectively.}
\label{Qspectrum}
\end{figure}

 The spectrum of the operator $- {\hat Q}_{[X,T]} / k $
is shown at Fig. \ref{Qspectrum} and we are interested here
in the particular functions $\xi_{1} (\theta )$ and $\lambda_{1}$.

 It's not difficult to see then that the eigen-function
$\Phi_{\theta} (\theta)$ and $\xi_{1} (\theta )$ correspond
to the gap edges $E_{3} = 0$ and $E_{4} = - \lambda_{1} / k$.
Easy to see also that in the limit of the small amplitude of
oscillations we have $|E_{4} - E_{3}| \rightarrow 0$ in the
full accordance with the perturbations theory. The expressions
for the functions $\xi_{1} (\theta )$ and $\lambda_{1}$ can be
written in the form (see \cite{CoopKharSukh}):

$$\xi_{1} (\theta, a, s, \gamma ) \,\, \sim \,\,
{\rm dn} \, \left({K (s) \over \pi} \theta, s \right)
\left[ 1 \, + \, 2 s^{2} \, - \, \sqrt{1 - s^{2} + 4 s^{4}} 
\, - \, 5 s^{2} \, {\rm sn}^{2} 
\left({K (s) \over \pi} \theta, s \right) \right] $$
$$\lambda_{1} (a, s, \gamma ) \,\, = \,\, - \, \left. K^{2} (s)
\left( 2 \sqrt{1 - s^{2} + 4 s^{4}} \, - \, 2 \, + \, s^{2} 
\right) \, k^{3} \, \right/ \, \pi^{2} \, = $$
$$= \, - \, \left. \pi \left( 2 \sqrt{1 - s^{2} + 4 s^{4}} 
\, - \, 2 \, + \, s^{2} \right) 
\left( {a \over 6 s^{2}}\right)^{3/2} \right/ K (s) $$
in our notations.

 By direct substitution it is not difficult to also  get
the following relations for the values of $\omega (k, A, n)$,
$\Phi$, ${\hat Q}$, $\xi_{1}$ and $\lambda_{1}$:

\begin{equation}
\label{appr1}
\omega \, = \, - \, k \, n \, + \, k^{3} \, - \,
{a_{0}^{2} \over 24 k} \, + \, {\cal O} (a_{0}^{4}) 
\end{equation}  
\begin{equation}
\label{appr2}
\Phi (\theta, k, A, n) \, = \, n \, + \, a_{0} \, \cos \theta
\, + \, {a_{0}^{2} \over 12 k^{2}} \cos 2\theta \, + \,
{\cal O} (a_{0}^{3}) 
\end{equation}
\begin{equation}
\label{appr3}
{\hat Q}_{[k,A,n]} \, = \, k^{3} \, - \, {a_{0}^{2} \over 24 k}
\, + \, k \, a_{0} \, \cos \theta \, + \, 
{a_{0}^{2} \over 12 k} \cos 2\theta \, + \,
k^{3} {\partial^{2} \over \partial \theta^{2}} \, + \,
{\cal O} (a_{0}^{3}) 
\end{equation}
\begin{equation}
\label{appr4}
\xi_{1} (\theta, k, A, n) \, = \, \cos \theta \, - \,
{a_{0} \over 2 k^{2}} \, + \, {a_{0} \over 6 k^{2}} \,
\cos 2\theta \, + \, {\cal O} (a_{0}^{2}) 
\end{equation}
\begin{equation}
\label{appr5}
\lambda_{1} \, = \, - {5 a_{0}^{2} \over 12 k} \, + \,
{\cal O} (a_{0}^{4}) 
\end{equation}

 We can see that the solutions (\ref{Philsol}) become singular
in the limit of the small amplitude oscillations $A \rightarrow 0$
if we do not put additional requirement
$$\langle \xi_{1} , g_{(l)} \rangle \, \equiv \, 0 $$
for all $g_{(l)}$. 

 The idea of correction of the Whitham approach for the almost
linear case using a correction of a dispersion relation
was invented by A.C. Newell (see \cite{Newell}, Chapter 2).
In \cite{DefLin} the method of A.C. Newell was generalized to
the case of the deformed Whitham systems which gives the regular
deformation procedure in the limit of the small amplitude of
oscillations.

 To improve the deformation procedure described above we will use 
the deformation scheme suggested in \cite{DefLin} for the case of 
almost linear systems.

 Namely, the orthogonality of $g_{(l)}(\theta, X, T)$ to
$\xi_{1} (\theta, X, T)$ can be provided in the following way:

 First of all we choose the parameters $(k, A, n)$ instead of
$(k, \omega, n)$ as the regular parameters everywhere (including
the region $A \rightarrow 0$). Now the main approximation in the
asymptotic solution (\ref{kdvexp}) will be again given by the
function $\Phi (S(X,T) + \theta, \, k , A , n)$ such that
$S_{X}(X,T) = k (X,T)$. So we have again the same approximation
with the same relation between $S$ and $k$ as previously at 
every $T$. However, we make now also the "deformation" of
time evolution of phase $S(X,T)$ such that 
$S_{T} (X,T) \neq \omega (k, A, n)$ anymore. Instead, we put
now the deformed relation
\begin{equation}
\label{defdisp}
S_{T} \,\, = \,\, \omega (k, A, n) \, + \,
\sum_{l\geq1} \omega_{(l)} (k, A, n, k_{X}, A_{X}, n_{X}, \dots )
\end{equation}
connecting the time derivative $S_{T}$ and the parameters
$(k, A, n)$ of the main approximation. Here again all the
functions $\omega_{(l)} (k, A, n, k_{X}, A_{X}, n_{X}, \dots )$
are differential polynomials in 
$(k_{X}, A_{X}, n_{X}, \dots )$ of the degree $l$ with coefficients  
smooth in $(k, A, n)$ according to 
the same gradation rule, i.e.

all the functions $f (k, A, n)$ have degree $0$;

the derivatives $k_{lX}$, $A_{lX}$, $n_{lX}$
have degree $l$;

the degree of the product of homogeneous differential polynomials is 
equal to the sum of their degrees.

As we have already said the parameter $A = \Phi_{max} - \Phi_{min}$
plays here the role of the amplitude of oscillations and we
have $A (X,T) \sim a_{0} (X,T)$ for the small $A$.

We write now the deformed Whitham system in the form
\begin{eqnarray}
&&
k_{T} \,\, = \,\, \left( \omega (k, A, n) \, + \,
\sum_{l\geq1} \omega_{(l)} (k, A, n, k_{X}, A_{X}, n_{X}, \dots )
\right)_{X} 
\nonumber\\
&&
\label{kansyst}
A_{T} \,\, = \,\, 
\sum_{l\geq1} \alpha_{(l)} (k, A, n, k_{X}, A_{X}, n_{X}, \dots )
\\
&&
n_{T} \,\, = \,\,
\sum_{l\geq1} \eta_{(l)} (k, A, n, k_{X}, A_{X}, n_{X}, \dots )
\nonumber
\end{eqnarray}
which gives a full deformation of the Whitham system having a
regular behavior in the case of small amplitudes.

 The functions $\alpha_{(l)}$, $\eta_{(l)}$ are defined as 
previously from the orthogonality conditions of the functions
$f_{(l)}(\theta, X, T)$ to the ``left" eigen-vectors $\Phi (\theta)$
and $1$ of the operator ${\hat L}$ corresponding to zero 
eigenvalues. The functions $\omega_{(l)}$ in (\ref{defdisp})
are defined now from the orthogonality of the functions
$g_{(l)}(\theta, X, T)$ to the eigen-vector
$\xi_{1} (\theta, X, T)$ of the operator ${\hat Q}_{[X,T]}$ 
corresponding to the ``small" eigenvalue 
$\lambda_{1} (k, A, n)$. 

So now we have the condition
\begin{equation}
\label{xiortcond}
\int_{0}^{2\pi} \xi_{1} (\theta, X, T) \, 
g_{(l)}(\theta, X, T) \, {d \theta \over 2\pi} 
\,\, \equiv \,\, 0
\end{equation}
in addition to conditions (\ref{ortcond}). The functions
$\lambda_{1} (k, A, n)$, $\xi_{1} (\theta, k, A, n)$ are
defined by continuity on the whole family of one-phase solutions
so we can define the system (\ref{kansyst}) on the full space of
parameters.

For our choice of the functions $\Phi (\theta, k, A, n)$ 
it is easy to prove that
the function $\xi_{1} (\theta, k, A, n)$ is even in
$\theta$.

For the solutions $\Phi_{(l)} (\theta, X, T)$ we will have
automatically
\begin{equation}
\label{xinormcond}
\int_{0}^{2\pi} \xi_{1} (\theta, X, T) \,
\Phi_{(l)} (\theta, X, T) \, {d \theta \over 2\pi} 
\,\, \equiv \,\, 0
\end{equation}
in addition to normalization conditions (\ref{normcond}).

In the same way as previously the following lemma can be proved
for systems (\ref{defdisp})-(\ref{kansyst}) and the
asymptotic expansion
\begin{equation}
\label{phiexp}
\phi (\theta, X, T) \,\, = \,\,
\Phi (S(X,T) + \theta, k, A, n) \, + \,
\sum_{l\geq1} \Phi_{(l)} (S(X,T) + \theta, X, T) 
\end{equation}

\vspace{0.5cm}   

{\bf Lemma 3.1.}
 {\it For the "unified" choice of the functions
$\Phi(\theta, k, A, n)$ corresponding to
Fig. \ref{zerophase} the following statements are true: 

1) All even terms $\sigma_{(2l)}(k, A, n, \dots)$,
$\eta_{(2l)}(k, A, n, \dots)$
in the deformation of the Whitham system (\ref{kansyst}) are
identically zero: $\alpha_{(2l)} \equiv 0$,
$\eta_{(2l)} \equiv 0$;
 
2) All odd terms $\omega_{(2l+1)} (k, A, n, \dots )$,
$l \geq 0$ in the deformation (\ref{defdisp}) of the dispersion 
relation are identically zero: $\omega_{(2l+1)} \equiv 0$; 

3) All odd corrections $\Phi_{(2l+1)}(\theta,X,T)$,
$l \geq 0$ in (\ref{kdvexp}) are odd in $\theta$;

4) All even corrections $\Phi_{(2l)}(\theta,X,T)$,
$l \geq 1$ in (\ref{kdvexp}) are even in $\theta$. }

\vspace{0.5cm}

So we can rewrite the relation (\ref{defdisp}) and the system
(\ref{kansyst}) in the form 
\begin{eqnarray}
&&
\label{defdisp1} 
S_{T} \,\, = \,\, \omega (k, A, n) \, + \,
\sum_{l\geq1} \omega_{(2l)} (k, A, n, k_{X}, A_{X}, n_{X}, \dots )
\nonumber\\
&&
k_{T} \,\, = \,\, \left( \omega (k, A, n) \, + \,
\sum_{l\geq1} \omega_{(2l)} (k, A, n, k_{X}, A_{X}, n_{X}, \dots )
\right)_{X}
\nonumber\\
&&\label{kansyst1}
\\
&&
A_{T} \,\, = \,\,
\sum_{l\geq0} \alpha_{(2l+1)} (k, A, n, k_{X}, A_{X}, n_{X}, \dots )
\nonumber\\
&&
n_{T} \,\, = \,\,
\sum_{l\geq0} \eta_{(2l+1)} (k, A, n, k_{X}, A_{X}, n_{X}, \dots ).
\nonumber
\end{eqnarray}
We can see that for our choice of the functions 
$\Phi (\theta; k, A, n)$ the full deformation (\ref{kansyst1})
of the Whitham system includes only odd degrees of the expansion
in higher derivatives which emphasizes the dispersive character
of the deformation.

\vspace{0.5cm}

To calculate the terms $\omega_{(2)}$, $\alpha_{(3)}$,
$\eta_{(3)}$ let us write down the expressions for the
discrepancies $f_{(1)}$, $f_{(2)}$, $f_{(3)}$ in the form
\begin{eqnarray}
&&
- f_{(1)} \, = \, \Phi_{T}^{[1]} + \Phi \Phi_{X} +
3 S_{X} S_{XX} \Phi_{\theta\theta} + 3 S_{X}^{2} \Phi_{\theta\theta X}
+ S_{T}^{[1]} \Phi_{\theta}
\nonumber\\
&&
\nonumber\\
&&
- f_{(2)} \, = \, \Phi_{T}^{[2]} + \Phi_{(1) T}^{[2]} +
3 S_{XX} \Phi_{\theta X} + 3 S_{X} \Phi_{\theta XX} + 
S_{XXX} \Phi_{\theta} + S_{X} \Phi_{(1)} \Phi_{(1) \theta} + 
\nonumber\\
&&
\nonumber\\
&&
+ \Phi  \Phi_{(1) X} +  \Phi_{(1)} \Phi_{X} + 
3 S_{X} S_{XX} \Phi_{(1) \theta\theta} +
3 S_{X}^{2} \Phi_{(1) \theta\theta X} + S_{T}^{[2]} \Phi_{\theta} 
\nonumber\\
&&
\nonumber\\
&&
- f_{(3)} \, = \, \Phi_{T}^{[3]} + \Phi_{(1) T}^{[3]} + 
\Phi_{(2) T}^{[3]} + \Phi_{XXX} + 3 S_{XX} \Phi_{(1) \theta X} +
3 S_{X} \Phi_{(1) \theta XX} + S_{XXX} \Phi_{(1) \theta} +
\Phi_{(1)} \Phi_{(1) X}
\nonumber\\
&&
+ \Phi  \Phi_{(2) X} + \Phi_{(2)} \Phi_{X} + 
3 S_{X} S_{XX} \Phi_{(2) \theta\theta} +
3 S_{X}^{2} \Phi_{(2) \theta\theta X} + 
S_{X} \Phi_{(1)} \Phi_{(2) \theta} +
S_{X} \Phi_{(2)} \Phi_{(1) \theta} +
S_{T}^{[3]} \Phi_{\theta}.
\nonumber
\end{eqnarray}
Let us remind again that we do not prescribe any certain degree to 
the operator $\partial/\partial T$, so we have
\begin{eqnarray}
&& 
S_{T} \, = \, \omega (k, A, n) \, + \,
\omega_{(1)} (k, A, n, k_{X}, A_{X}, n_{X}) \, + \, \dots 
\nonumber\\
&&
\Phi_{T} \, = \, \Phi_{T}^{[1]} \, + \, \Phi_{T}^{[2]} \, + \,
\Phi_{T}^{[3]} \, + \, \dots 
\nonumber\\
&&
\Phi_{(1) T} \, = \, \Phi_{(1) T}^{[2]} \, + \, \Phi_{(1) T}^{[3]} 
\, + \, \Phi_{(1) T}^{[4]} \, + \, \dots 
\nonumber\\
&&
\Phi_{(2) T} \, = \, \Phi_{(2) T}^{[3]} \, + \, \Phi_{(2) T}^{[4]}
\, + \, \Phi_{(2) T}^{[5]} \, + \, \dots 
\nonumber
\end{eqnarray}
where all $\Phi_{(l) T}^{[s]}$ are differential polynomials in 
$(k_{X}, A_{X}, n_{X}, \dots )$ of the degree $s$ with smooth 
coefficients depending on $(k, A, n)$.

It is easy to see that the only odd in $\theta$ term in
$f_{(1)}$ is $- S_{T}^{[1]} \Phi_{\theta}$. So from the
orthogonality of $g_{(1)}$ to $\xi_{1} (\theta, X, T)$ we get
immediately $\omega_{(1)} \equiv S_{T}^{[1]} \equiv 0$ in
accordance to Lemma 2. In the same way we put also 
$\omega_{(l)} \equiv 0$, $\delta_{(l)} \equiv 0$ for all
$l = 2s + 1$, $s \geq 0$, such that only 
$\omega_{(2s)}$, $\delta_{(2s)}$ should be computed for
$s \geq 1$.

 We will not need to calculate completely the system (\ref{kansyst1})
in this paper, however, let us briefly describe here the scheme
for the determination of the functions $\alpha_{(1)}$, $\eta_{(1)}$,
$\alpha_{(3)}$, $\eta_{(3)}$, $\omega_{(2)}$. We have 
$$\Phi_{T}^{[1]} \, = \, \Phi_{a} A_{T}^{[1]} +
\Phi_{n} n_{T}^{[1]} + \Phi_{k} k_{T}^{[1]} \, = \,
\Phi_{a} \alpha_{(1)} + \Phi_{n} \eta_{(1)} +
\Phi_{k} \left( \omega(k,a,n) \right)_{X} $$
so the orthogonality of $f_{(1)} (\theta, X, T)$ to the
functions $\Phi (\theta, X, T)$ and $1$ gives the usual
expression for $\alpha_{(1)}$, $\eta_{(1)}$ given by the standard
system of Whitham.
The only  even term in $f_{(2)}$ is $- \Phi_{T}^{[2]}$.
So we get immediately $\alpha_{(2)} \equiv 0$,
$\eta_{(2)} \equiv 0$ from the orthogonality of $f_{(2)}$ to
$\Phi (\theta, X, T)$ and $1$. The term $- \Phi_{(1) T}^{[2]}$
is given by
$$- \int \left( {\delta \Phi_{(1)}(X) \over \delta A(Y)}
\, \alpha_{(1)}(Y) \, + \, {\delta \Phi_{(1)}(X) \over \delta n(Y)}
\, \eta_{(1)}(Y) \, + \, {\delta \Phi_{(1)}(X) \over \delta k(Y)}
\, \omega_{Y}(Y) \right) \, dY $$
and it is a known function. From the orthogonality of $g_{(2)}$ to
$\xi_{1} (\theta, X, T)$ we get a relation for
$\omega_{(2)} (k, A, n, \dots)$:
$$\omega_{(2)} \, \int_{0}^{2\pi} \xi_{1} (\theta) \,
\left( \Phi (\theta) - \Phi (0) \right) 
\, {d \theta \over 2\pi} \,\, = \,\,
\int_{0}^{2\pi} \xi_{1} (\theta) \, g^{\prime}_{(2)} (\theta) 
\, {d \theta \over 2\pi} $$
where
$$g^{\prime}_{(2)} \, = \, - \int_{0}^{\theta} \left[ 
\Phi_{(1) T}^{[2]} + 3 k_{X} \Phi_{\theta^{\prime} X}
+ 3 k \Phi_{\theta^{\prime} XX} + k_{XX} \Phi_{\theta^{\prime}} + 
k \Phi_{(1)} \Phi_{(1) \theta^{\prime}} + \Phi \Phi_{(1) X} + 
\right. $$
$$\left. + \Phi_{(1)} \Phi_{X} + 
3 k k_{X} \Phi_{(1) \theta^{\prime}\theta^{\prime}} +
3 k^{2} \Phi_{(1) \theta^{\prime}\theta^{\prime} X} \right] 
\, d \theta^{\prime} \,\, + \,\, \delta_{(2)} (X,T) $$

It convenient to determine the values of $\omega_{(2)}$
and $\delta_{(2)}$ simultaneously from the orthogonality
of $g_{(2)} (\theta)$ to both the vectors $\xi_{1} (\theta)$
and 
$$\xi_{tot} (\theta) \,\, = \,\, \sum_{s \geq 1} 
{1 \over \lambda_{2s}} \, J_{2s} \, \xi_{2s} \,\,\,\,\, ,
\,\,\,\,\, J_{2s} \,\, = \,\, \int_{0}^{2\pi} \xi_{2s} (\theta)
\, {d \theta \over 2\pi} $$
and add the relation
$$\omega_{(2)} \, \int_{0}^{2\pi} \xi_{tot} (\theta) \,
\left( \Phi (\theta) - \Phi (0) \right)
\, {d \theta \over 2\pi} \,\, = \,\,
\int_{0}^{2\pi} \xi_{tot} (\theta) \, g^{\prime}_{(2)} (\theta)  
\, {d \theta \over 2\pi} $$
which gives a non-degenerate linear system on the values
$\omega_{(2)}$, $\delta_{(2)}$.\footnote{Both the functions
$\xi_{1} (\theta)$, $\xi_{tot} (\theta)$ have in fact explicit
expressions in terms of elliptic functions.}

Repeating all the arguments we get 
$\omega_{(3)} \equiv S_{T}^{[3]} \equiv 0$ from the orthogonality
of $g_{(3)}$ to $\xi_{1} (\theta, X, T)$. We have also 
$\Phi_{(1) T}^{[3]} \equiv 0$ view $A_{T}^{[2]} \equiv 0$,
$n_{T}^{[2]} \equiv 0$, $k_{T}^{[2]} \equiv 0$. The function
$- \Phi_{(2) T}^{[3]}$ is given by
$$- \int \left( {\delta \Phi_{(2)}(X) \over \delta A(Y)}
\, \alpha_{(1)}(Y) \, + \, {\delta \Phi_{(2)}(X) \over \delta n(Y)}
\, \eta_{(1)}(Y) \, + \, {\delta \Phi_{(2)}(X) \over \delta k(Y)}
\, \omega_{Y}(Y) \right) \, dY $$
and is a known function again.

We have then
$$\Phi_{T}^{[3]} \, = \, 
\Phi_{A} \alpha_{(3)} + \Phi_{n} \eta_{(3)} +
\Phi_{k} \left( \omega_{(2)} \right)_{X} $$
so we get the functions $\alpha_{(3)}$, $\eta_{(3)}$ from the
orthogonality of $f_{(3)}$ to $\Phi (\theta, X, T)$ and $1$.

It is easy to see also that the procedure can be extended to any order
$l$ such that all the terms $\omega_{(2l)}$, $\alpha_{(2l+1)}$,
$\eta_{(2l+1)}$ will be uniquely determined.

\vspace{0.5cm}

The system (\ref{kansyst1}) determines the
evolution of the parameters $(k, A, n)$ of the zero approximation of
(\ref{kdvexp}) such that the following conditions are satisfied:

I$^{\prime}$ ) All the functions $\Phi (\theta; k, A, n)$ are 
chosen in the way shown at Fig. 1.

II$^{\prime}$ ) The modulated phase $S (X,T)$ and the
parameters $(k, A, n)$ of the zero approximation are connected 
by the relation

$$ S_{X} (X,T) \, = \, k (X,T). $$

III$^{\prime}$ ) The higher corrections 
$\Phi_{(l)} (\theta, X, T)$, $l \geq 1$ 
satisfy normalization conditions (\ref{normcond}) 
and (\ref{xinormcond}).

\vspace{0.5cm}

 We would like to introduce a small parameter $\epsilon$ 
according to our gradation rule for more convenient notations. 
System (\ref{kansyst1}) will be rewritten 
in the form
\begin{eqnarray}
\label{epsdisp}
&&
S_{T} \,\, = \,\, \omega (k, A, n) \, + \,
\sum_{l\geq1} \epsilon^{2l} \,
\omega_{(2l)} (k, A, n, k_{X}, A_{X}, n_{X}, \dots )
\\
&&
k_{T} \,\, = \,\, \left( \omega (k, A, n) \, + \,
\sum_{l\geq1} \epsilon^{2l} \,
\omega_{(2l)} (k, A, n, k_{X}, A_{X}, n_{X}, \dots )
\right)_{X}
\nonumber\\
&&
\label{epskan}
A_{T} \,\, = \,\,
\sum_{l\geq0} \epsilon^{2l} \,
\alpha_{(2l+1)} (k, A, n, k_{X}, A_{X}, n_{X}, \dots )
\\
&&
n_{T} \,\, = \,\,
\sum_{l\geq0} \epsilon^{2l} \,
\eta_{(2l+1)} (k, A, n, k_{X}, A_{X}, n_{X}, \dots ).
\nonumber
\end{eqnarray}
The asymptotic expansion (\ref{phiexp}) will also be rewritten
in the form
\begin{equation}
\label{phiepsexp}
\phi (\theta, X, T) \,\, = \,\,
\Phi \left( {S(X,T) \over \epsilon} + \theta, k, A, n \right)
\, + \, \sum_{l\geq1} \epsilon^{l} \, \Phi_{(l)}
\left( {S(X,T) \over \epsilon} + \theta, X, T \right)
\end{equation}
according to the gradation rules for the functions $S(X,T)$ and
$\Phi_{(l)}(\theta, X, T)$.

In these new notations we can actually see that system
(\ref{e2disp})-(\ref{e2kan}) describes the asymptotic solutions
of the equation
\begin{equation}
\label{ekdv}
\varphi_{T} \, + \, \varphi \, \varphi_{X} \, + \, 
\epsilon^{2} \, \varphi_{XXX} \, = \, 0
\end{equation}
where the small dispersion $\epsilon^{2}$ arises after the
rescaling $T \rightarrow \epsilon T$,
$X \rightarrow \epsilon X$. So our further considerations will 
be applied to the KdV equation in the small-dispersion form 
(\ref{ekdv}).

 Let us emphasize, however, that (\ref{phiepsexp}) is not
an $\epsilon$-expansion of the asymptotic solution of (\ref{ekdv})
since all of the functions 
$k (X, T, \epsilon)$, $A (X, T, \epsilon)$,
$n (X, T, \epsilon)$ are solutions of  
$\epsilon$-dependent system (\ref{epskan}), such that 
expansion (\ref{phiepsexp}) can contain more complicated 
$\epsilon$-dependence. According to our rules we should not separate 
the different orders in $\epsilon$ of the functions
$k (X, T, \epsilon)$, $A (X, T, \epsilon)$,
$n (X, T, \epsilon)$ and just use the gradation rules
formulated above for the $\epsilon$-dependent functions. 
\footnote{As it was pointed out in
\cite{deform1} the series (\ref{phiepsexp}) (or (\ref{phiexp}))
corresponds to the expansion with respect to 
$X$-derivatives of the ``renormalized" $\epsilon$-dependent 
parameters of the main approximation 
$\Phi(\theta, k, A, n)$ which gives these specific rules of
constructing of the series (\ref{phiepsexp}).}

 The solutions of system (\ref{epskan}) can be considered in 
different ways. Thus, it is easy to define the formal graded 
form of solutions of (\ref{epskan})

$$k (X,T) \, \, = \,\, k (X,0) \, + \, \sum_{l\geq 1}
T^{l} \, \epsilon^{l-1} \, K_{(l)} \left( k (X,0), A (X,0),
n (X,0), \dots \right) $$
$$A (X,T) \, \, = \,\, A (X,0) \, + \, \sum_{l\geq 1}
T^{l} \, \epsilon^{l-1} \, A_{(l)} \left( k (X,0), A (X,0),
n (X,0), \dots \right) $$
$$n (X,T) \, \, = \,\, n (X,0) \, + \, \sum_{l\geq 1}
T^{l} \, \epsilon^{l-1} \, N_{(l)} \left( k (X,0), A (X,0),
n (X,0), \dots \right) $$
for $0 < T < \delta$, where all $K_{(l)}$, $A_{(l)}$, $N_{(l)}$
are local functionals of $k (X,0)$, $A (X,0)$, $n (X,0)$ and their
derivatives having the corresponding degree.

 However, a more complicated treatment of the solutions of 
(\ref{epskan}) connected with their global behavior based on
the so-called quasitriviality transformations 
(see \cite{DubrZhang1,DubrZhang2}) of parameters 
$(k, A, n)$ is also possible and seems to be very important
in the theory of the deformed Whitham systems.

 Let us recall also that the KdV equation (\ref{kdv}) has an
infinite series of conservation laws which can be written in the
form

\begin{equation}
\label{nuconserv}
{d \over d t} \, {\cal P}^{\nu} (\varphi, \varphi_{x}, \dots )
\,\,\, = \,\,\,
{d \over d x} \, {\cal Q}^{\nu} (\varphi, \varphi_{x}, \dots )
\end{equation}
with some local functionals 
${\cal P}^{\nu} (\varphi, \varphi_{x}, \dots )$,
${\cal Q}^{\nu} (\varphi, \varphi_{x}, \dots )$. For equation
(\ref{ekdv}) the corresponding relations can be written 
respectively

\begin{equation}
\label{epsnuconserv}
{d \over d T} \, {\cal P}^{\nu} 
(\varphi, \epsilon \varphi_{X}, \dots )
\,\,\, = \,\,\,
{d \over d X} \, {\cal Q}^{\nu} 
(\varphi, \epsilon \varphi_{X}, \dots ) \,\,\,\,\, ,
\,\,\,\,\, \nu = 0, 1, 2, \dots
\end{equation}

 According to standard numeration we put

$${\cal P}^{0} \,\, = \,\, \varphi \,\,\,\,\, , \,\,\,\,\, 
{\cal Q}^{0} \,\, = \,\, - \, {1 \over 2} \, \varphi^{2} \, - \,
\epsilon^{2} \varphi_{XX} $$
for $\nu = 0$ and we have the conservation of the Casimir function

$$ N \,\, = \,\, \int_{-\infty}^{+\infty} \varphi \, d X $$
for the Gardner - Zakharov - Faddeev bracket in this case.

 For $\nu = 1$ it is put traditionally

$${\cal P}^{1} \,\, = \,\, {1 \over 2} \, \varphi^{2} 
\,\,\,\,\, , \,\,\,\,\, {\cal Q}^{1} \,\, = \,\,
- \, {1 \over 3} \, \varphi^{3}
\, - \, \epsilon^{2} \varphi \varphi_{XX}
\, + \, {1 \over 2} \, \epsilon^{2} \varphi_{X}^{2} $$
which corresponds to the conservation of the momentum functional
for the same bracket.

 For $\nu = 2$ we put

$${\cal P}^{2} \, = \, {1 \over 6} \, \varphi^{3}
\, - \, {1 \over 2} \, \epsilon^{2} \varphi_{X}^{2} 
\,\,\, , \,\,\,
{\cal Q}^{2} \, = \, - \, {1 \over 8} \, \varphi^{4}
\, - \, {1 \over 2} \, \epsilon^{2} \varphi^{2} \varphi_{XX}
\, + \, \epsilon^{2} \varphi \varphi_{X}^{2} \, + \,
\epsilon^{4} \varphi_{X} \varphi_{XXX} \, - \,
{1 \over 2} \,  \epsilon^{4} \varphi_{XX}^{2} $$
which gives the conservation of energy in the 
Gardner - Zakharov - Faddeev Poisson structure.

 The higher conservation laws are connected with the integrable
nature of the KdV equation and arise from the method of the
inverse scattering problem.

 It's not difficult to see that the conservation laws of the
KdV equation (\ref{ekdv}) give conservation laws for system
(\ref{epskan}) after the "averaging" of the corresponding
densities ${\cal P}^{\nu}$, ${\cal Q}^{\nu}$ on the asymptotic
family (\ref{phiepsexp}). Indeed, after the substitution of
(\ref{phiepsexp}) into (\ref{epsnuconserv}) and integration
w.r.t. $\theta$ we get the relations

\begin{equation}
\label{nuavlaws}
{d \over d T} \, \langle {\cal P}^{\nu} \rangle \,\,\, =
\,\,\, {d \over d X} \, \langle {\cal Q}^{\nu} \rangle
\end{equation}
where the quantities $<{\cal P}^{\nu}>$, $<{\cal Q}^{\nu}>$
are given by the substitution of solutions (\ref{phiepsexp})
in the expressions for ${\cal P}^{\nu}$ and ${\cal Q}^{\nu}$
and integration w.r.t. $\theta$ over the period. It's not
difficult to see also that the values $<{\cal P}^{\nu}>$,
$<{\cal Q}^{\nu}>$ are expressed in this case as the local 
functionals of the parameters $(k, A, n)$ and their
$X$-derivatives

$$\langle {\cal P}^{\nu} \rangle \, = \, 
\langle {\cal P}^{\nu} \rangle 
\left( k, A, n, k_{X}, A_{X}, n_{X}, \dots \right) 
\,\,\, , \,\,\,
\langle {\cal Q}^{\nu} \rangle \, = \,      
\langle {\cal Q}^{\nu} \rangle
\left( k, A, n, k_{X}, A_{X}, n_{X}, \dots \right) $$
which are polynomial in the derivatives of $(k, A, n)$ and
can be written in the graded form we introduced above.
Relations (\ref{nuavlaws}) give then an infinite set of
conservation laws for system (\ref{epskan}) written in the 
same graded form.

 We can see also that the values of independent integrals
$<{\cal P}^{\nu}>$ can be also chosen as the parameters of the
solutions (\ref{phiepsexp}) such that the values $(k, A, n)$
will be expressed in the form of graded expansions with
respect to the $X$-derivatives of $<{\cal P}^{\nu}>$ given
by the "inversion" of the corresponding expansions for
$<{\cal P}^{\nu}>$. System (\ref{epskan}) written in the 
corresponding parameters (say $<{\cal P}^{0}>$, 
$<{\cal P}^{1}>$, $<{\cal P}^{2}>$) has then a conservative
form and expresses the balance of the chosen conservation laws.

\vspace{0.5cm}

 {\bf Remark 3.1.}

 Let us come back now to Fig. \ref{Qspectrum} representing the 
spectrum of the operator $ - {\hat Q}_{[X,T]}/k$. Let us consider 
the limit $k \rightarrow 0$ now and consider the spectrum of
$ - {\hat Q}_{[X,T]}/k$ on the space of $2\pi$-periodic functions
in this limit. We can say first af all that the sizes of the
energy bands $[E_{0}, E_{1}]$, $[E_{2}, E_{3}]$, $[E_{4}, E_{5}]$
tend to zero in this situation giving the "splitting" of the
three localized quantum states of the corresponding decreasing
(one-soliton) potential arising at $k \rightarrow 0$.\footnote{The
limit $k \rightarrow 0$ of a one-phase solution of KdV gives a
one-soliton solution corresponding to a reflectionless 
potential with one localized quantum state for the Lax operator
(\ref{ShrOp}). The same solution gives a potential with three
bounded states (one with $E = 0$) for the operator 
${\hat Q}_{[X,T]}$ given by (\ref{Qoper}) in the same limit.} 
The potential $\Phi (k x ; k, A, n)$ represents in this case 
a "lattice" of distant one-soliton solutions with the period 
$\sim k^{-1}$ with $k \rightarrow 0$.

 Let us note now that the eigen-values of the operator
$ - {\hat Q}_{[X,T]}/k$ for the energies $E > E_{6}$ are
double-degenerated on the space of periodic functions and
represent the "boundaries of the gaps of zero width" in the
spectrum of $ - {\hat Q}_{[X,T]}/k$. It's not difficult to see
also that the distance between these eigen-values decreases
($\sim k^{2}$) in the limit $k \rightarrow 0$. Moreover, the 
size of the gap $[E_{5}, E_{6}]$ (as well as the energy band
$[E_{4}, E_{5}]$) decreases also in this situation. As a result,
we can see that another instability arises in our scheme for
the case of the small $k$ due to the large number of the "small"
eigen-values of ${\hat Q}_{[X,T]}$ in this limit. This fact
means most probably that the averaging methods are not very
applicable in the limit $k \rightarrow 0$ where the 
"multi-soliton" description seems to give more adequate
picture.

\vspace{0.5cm}

 At last let us say that we believe that it's enough to keep just 
the first ($\epsilon^{2}$) dispersive terms in system (\ref{epskan})
for the description of many oscillating regimes arising in the
KdV theory. Finally we arrive at the system

\begin{equation}
\label{e2disp}
S_{T} \,\, = \,\, \omega (k, A, n) \, + \, \epsilon^{2} \,
\omega_{(2)} (k, A, n, \dots) 
\end{equation}
\begin{eqnarray}
&&
k_{T} \,\, = \,\, \left(
\omega (k, A, n) \, + \, \epsilon^{2} \,
\omega_{(2)} (k, A, n, \dots) \right)_{X} 
\nonumber\\
&&
\label{e2kan}
A_{T} \,\, = \,\, a_{(1)} (k, A, n, k_{X}, A_{X}, n_{X}) 
\, + \, \epsilon^{2} \, a_{(3)} (k, A, n, \dots)
\\
&&
n_{T} \,\, = \,\, \eta_{(1)} (k, A, n, k_{X}, A_{X}, n_{X})
\, + \, \epsilon^{2} \, \eta_{(3)} (k, A, n, \dots) 
\nonumber
\end{eqnarray}
since we believe that it demonstrates already many essential
features of the full system (\ref{epsdisp})-(\ref{epskan}).

 System (\ref{e2kan}) should be considered as a system of differential
equations in the ordinary sense, in particular, all the solutions
of (\ref{e2kan}) are supposed to be well defined functions of $X$
and $T$ with some concrete behavior depending on the regime
under investigation.

 In the next chapters we are going to consider the questions 
connected with the Hamiltonian structures of system (\ref{epskan})
which is the main subject of this paper. So, we will consider now
the initial KdV equation as a part of an integrable hierarchy
having two local Hamiltonian structures and discuss a possibility
of the "averaging" of the Hamiltonian structures to obtain the
Hamiltonian structures of the Dubrovin - Zhang type for system
(\ref{epskan}).

\section{The commuting flows and the Hamiltonian structures.}
\setcounter{equation}{0}

 It is well known that the KdV equation represents the first 
nontrivial flow of the integrable KdV hierarchy generated by the
higher KdV integrals

$$I^{\nu} \,\, = \,\, \int_{-\infty}^{+\infty} {\cal P}^{\nu} 
(\varphi, \varphi_{x}, \dots ) \, dx $$
with respect to the Gardner - Zakharov - Faddeev bracket

$$\{ \varphi (x), \varphi (y) \} \,\, = \,\, 
\delta^{\prime} (x-y) $$
or the Magri bracket

$$\{ \varphi (x), \varphi (y) \} \,\, = \,\,
\delta^{\prime\prime\prime} (x-y) \, + \, 
{2 \over 3} \, \varphi (x) \, \delta^{\prime} (x-y) \, + \,
{1 \over 3} \, \varphi_{x} \, \delta (x-y) $$

 All the higher KdV flows have the similar form

\begin{equation}
\label{higherKdV}
\varphi_{t^{\nu}} \,\, = \,\, f^{\nu} (\varphi, \varphi_{x},
\varphi_{xx}, \dots )
\end{equation}
and give an infinite set of commuting integrable flows.

 The commuting flows (\ref{higherKdV}) can be also written in
the "small dispersion" form

\begin{equation}
\label{epshigherKdV}
\epsilon \, \varphi_{T^{\nu}} \,\, = \,\, 
f^{\nu} (\varphi, \epsilon \, \varphi_{X},
\epsilon^{2} \, \varphi_{XX}, \dots )
\end{equation}
which gives the commuting flows for the KdV equation written
in the form (\ref{ekdv}).

 It is natural to expect then that the higher flows 
(\ref{epshigherKdV}) of the KdV hierarchy generate the commuting
flows for the deformed Whitham system (\ref{epskan}) such that we 
get an "integrable" hierarchy starting from system (\ref{epskan})
on the "averaged" level. 

 We have to introduce now the "extended functional space" 
${\cal M} = \{\varphi (\theta, X)\}$ consisting of smooth functions
$\varphi (\theta, X)$ which are $2\pi$-periodic in $\theta$ at 
every $X$. For our further purposes we need to introduce also
a "submanifold" ${\cal K} \in {\cal M}$ corresponding to the set
of solutions (\ref{phiepsexp}) which will play the basic role in
our considerations. Let us note here that all our considerations
will be connected with the formal asymptotic series in the
derivatives of parameters of one-phase solutions of KdV so we
define also the submanifold ${\cal K}$ in the same form, i.e.
as a formal submanifold having the asymptotic sense.

 Thus, we define the submanifold ${\cal K}$ in the space
of functions $\varphi (\theta, X)$ by the following rule:

1) The function $\varphi (\theta, X)$ belongs to the family
${\cal K}$ if it represents one of solutions (\ref{phiepsexp}),
i.e.

$$\varphi (\theta, X) \,\, = \,\, 
\Phi \left( {S(X) \over \epsilon} + \theta, k, A, n \right)
\, + \, \sum_{l\geq1} \epsilon^{l} \, \Phi_{(l)}
\left( {S(X) \over \epsilon} + \theta, [k, A, n], X \right) $$
with some functions $( S(X), A(X), n(X) )$
where $k (X) = S_{X} (X)$;

2) We put the following relation between the functions $S (X)$
and $k (X)$:\footnote{Let us assume here that the relations
$k (X) \rightarrow 0$ for $X \rightarrow \pm \infty$ are
imposed. However, the procedure will give us a local deformed
Poisson bracket on the space $( k(X), A(X), n(X) )$, so this
condition will not be important in fact for the final result.}

\begin{equation}
\label{Skcond}
S (X) \,\, = \,\, {1 \over 2} \int_{-\infty}^{+\infty}
{\rm sgn} (X - Y) \, k (Y) \, d Y 
\end{equation}

 The functions $(k (X), A(X), n(X))$ play the role of 
"coordinates" on the submanifold ${\cal K}$, so we consider
${\cal K}$ a manifolds parametrized by three functional
parameters.

 Let us formulate here the Theorem which connects 
the higher flows (\ref{epshigherKdV}) with
the commuting flows of system (\ref{epskan}).

\vspace{0.5cm}

{\bf Theorem 4.1.}

{\it 
Every higher KdV flow (\ref{epshigherKdV}) leaves invariant the
family of formal solutions (\ref{phiepsexp}) and generates a
commuting flow for the deformed Whitham system (\ref{epskan})
which can be represented in the same graded form

\begin{eqnarray}
\label{dispTnu}
&&
S_{T^{\nu}} \,\, = \,\, \omega^{\nu} (k, A, n) \, + \,
\sum_{l\geq1} \epsilon^{2l} \,
\omega_{(2l)}^{\nu} (k, A, n, k_{X}, A_{X}, n_{X}, \dots ) 
\\
&&
k_{T^{\nu}} \,\, = \,\, \left( \omega^{\nu} (k, A, n) \, + \,
\sum_{l\geq1} \epsilon^{2l} \,
\omega_{(2l)}^{\nu} (k, A, n, k_{X}, A_{X}, n_{X}, \dots )
\right)_{X}
\nonumber\\
&&
\label{kAnTnu}
A_{T^{\nu}} \,\, = \,\,
\sum_{l\geq0} \epsilon^{2l} \,
\alpha_{(2l+1)}^{\nu} (k, A, n, k_{X}, A_{X}, n_{X}, \dots )
\\
&&
n_{T^{\nu}} \,\, = \,\,
\sum_{l\geq0} \epsilon^{2l} \,
\eta_{(2l+1)}^{\nu} (k, A, n, k_{X}, A_{X}, n_{X}, \dots ).
\nonumber
\end{eqnarray}
as system (\ref{epskan}).
}

\vspace{0.5cm}

Proof.

Let us consider the formal asymptotic series

\begin{equation}
\label{nuseries}
\varphi \left( \theta, X, T, T^{\nu} \right) \,\, = \,\,
\sum_{l\geq0} \epsilon^{l} \, \Psi_{(l)} \left(
{S (X, T) \over \epsilon} \, + \, \theta , X, T, T^{\nu} \right)
\end{equation}
where every function $\Psi_{(l)} ( \theta , X, T, T^{\nu} )$
is a local functional of $k_{0} (X,T) = S_{0X} (X,T)$, 
$A_{0} (X,T)$, $n_{0} (X,T)$ 
and their $X$-derivatives which is polynomial in
derivatives and has degree $l$ according to the gradation rule
we introduced above. We require that series (\ref{nuseries})
coincides with asymptotic series (\ref{phiepsexp}) with the
same parameters $k_{0} (X,T)$, $A_{0} (X,T)$, $n_{0} (X,T)$ 
for $T^{\nu} = 0$

$$\Psi_{(0)} ( \theta , X, T, 0 ) \,\, = \,\,
\Phi  ( \theta , k_{0}, A_{0}, n_{0}) \,\,\,\,\, , \,\,\,\,\,
\Psi_{(l)} ( \theta , X, T, 0 ) \,\, = \,\,
\Phi_{(l)} ( \theta , X, T) $$
and satisfies the higher KdV equation (\ref{epshigherKdV})
for $T^{\nu} > 0$.

 After the substitution of (\ref{nuseries}) into 
(\ref{epshigherKdV}) in the graded form we get a chain of 
evolution equations on the functions 
$\Psi_{(l)} ( \theta , X, T, T^{\nu} )$ at every degree $l$

\begin{equation}
\label{gradsystems}
{d \over d T^{\nu}} \Psi_{(l)} ( \theta , X, T, T^{\nu} )
\,\, = \,\, \Lambda_{(l)} \left( \bm{\Psi}, \bm{\Psi}_{\theta},
\bm{\Psi}_{X}, \dots , k_{0}, A_{0}, n_{0}, k_{0X}, A_{0X}, 
n_{0X}, \dots \right)
\end{equation}
where every $\Lambda_{(l)}$ depends only on $\Psi_{(s)}$
with $s \leq l$.

 It's not difficult to check the following relations for
$T^{\nu} = 0$

\begin{equation}
\label{PsiLambda}
\Psi_{(l)} (- \theta , X, T, 0 )  = 
(-1)^{l} \, \Psi_{(l)} ( \theta , X, T, 0 ) \,\,\, ,
\,\,\, \Lambda_{(l)} (- \theta , X, T, 0 )  = 
(-1)^{l+1} \, \Lambda_{(l)} ( \theta , X, T, 0 ) 
\end{equation}
$l \geq 0$.

 Let us assume for simplicity that all the systems 
(\ref{gradsystems}) have smooth solutions on some interval
$T^{\nu} \in [0, \delta]$ with our initial data such that we
get a unique formal series (\ref{nuseries}) satisfying our
requirements on the same interval. Since equation 
(\ref{epshigherKdV}) gives a commuting flow for the KdV 
equation (\ref{ekdv}) we get that series (\ref{nuseries}) 
gives a formal solution of (\ref{ekdv}) at every 
$T^{\nu} \in [0, \delta]$. However, series (\ref{nuseries})
can not be considered as the asymptotic series (\ref{phiepsexp})
for $T^{\nu} > 0$ since the normalization conditions 
(\ref{normcond}), (\ref{xinormcond}) will be in general destroyed
by the evolution systems (\ref{gradsystems}).\footnote{The function
$\Psi_{(0)} ( \theta , X, T, T^{\nu} )$ will remain the one-phase
solution for $T^{\nu} > 0$ in this situation, however, the
normalization $\Psi_{(0)\theta} (0 , X, T, T^{\nu}) = 0$
will be also destroyed by the higher KdV flow on the one-phase
solutions.}

 Nonetheless, series (\ref{nuseries}) can be represented in 
form (\ref{phiepsexp}) after a redefinition of parameters

\begin{equation}
\label{Tnuparchange}
\left( k_{0} (X,T),  A_{0} (X,T),  n_{0} (X,T) \right) 
\,\, \rightarrow \,\,
\left( k (X,T,T^{\nu}),  A (X,T,T^{\nu}),  n (X,T,T^{\nu}) \right)
\end{equation}
and a re-expansion of (\ref{nuseries}) in the new graded form.
It is convenient then to represent the redefinition of 
$(k, A, n)$ in the differential graded form (\ref{kAnTnu})
which gives the required evolution system on family 
(\ref{phiepsexp}). 

 Let us discuss finally the possibility of a construction of the 
required system (\ref{kAnTnu}) on the space of parameters 
$(k(X), A(X), n(X))$ which will prove the Theorem. Indeed, the
function $n (X, T, T^{\nu})$ is given by the integral

$$\int_{0}^{2\pi} \varphi (\theta, X, T, T^{\nu}) \,\, 
{d \theta \over 2 \pi} $$
according to the definition, so we get immediately the
graded equation

$${d n \over d T^{\nu}} \,\, = \,\, \sum_{l = 0}^{\infty}
\int_{0}^{2\pi} \Lambda_{(l)} (\theta, X, T, T^{\nu}) \,\,
{d \theta \over 2 \pi} $$
which makes satisfied the second relation (\ref{normcond})
for $T^{\nu} > 0$.

 Let us consider now the first relation (\ref{normcond}) and
relation (\ref{xinormcond}). We have to find now two more 
functions $S (X, T, T^{\nu})$, $A (X, T, T^{\nu})$ such that 
the function $\Phi (\theta, S_{X}, A, n)$ satisfies the 
conditions

$$\int_{0}^{2\pi} \Phi_{\theta} \left( 
{S (X, T, T^{\nu}) \over \epsilon} + \theta,  S_{X}, A, n
\right) \sum_{l \geq 0} \epsilon^{l} \, \Psi_{(l)} \left(
{S (X, T) \over \epsilon} + \theta, X, T, T^{\nu} \right) \,
{d \theta \over 2 \pi} \, \equiv \, 0 $$
$$\int_{0}^{2\pi} \xi_{1} \left(
{S (X, T, T^{\nu}) \over \epsilon} + \theta,  S_{X}, A, n
\right) \, \times $$
$$\times \, \left( \sum_{l \geq 0} \epsilon^{l} \, \Psi_{(l)} 
\left( {S (X, T) \over \epsilon} + \theta, X, T, T^{\nu} \right)
\, - \, \Phi \left( {S (X, T, T^{\nu}) \over \epsilon} + \theta,  
S_{X}, A, n \right) \right) \,
{d \theta \over 2 \pi} \, \equiv \, 0 $$
to be an appropriate main term in the "re-expanded" series
(\ref{nuseries}).

 We assume also $S (X, T, 0) = S_{0} (X, T)$, 
$k (X, T, 0) = k_{0} (X, T)$, $A (X, T, 0) = A_{0} (X, T)$,
$n (X, T, 0) = n_{0} (X, T)$ according to our scheme.

 Differentiating the first relation with respect to $T^{\nu}$
at $T^{\nu} = 0$ we get

$$\int_{0}^{2\pi} \left( {1 \over \epsilon} \, S_{T^{\nu}} \,
\Phi_{\theta\theta} \, + \, k_{T^{\nu}} \, \Phi_{\theta k}
\, + \, A_{T^{\nu}} \, \Phi_{\theta A} \, + \,
n_{T^{\nu}} \, \Phi_{\theta n} \right) \sum_{l \geq 0} 
\epsilon^{l} \, \Psi_{(l)} \, {d \theta \over 2 \pi} \, +$$
\begin{equation}
\label{Equation1}
+ \, \int_{0}^{2\pi} \Phi_{\theta} \sum_{l \geq 0} 
\epsilon^{l} \, \Lambda_{(l)}|_{(T^{\nu}=0)} \,
{d \theta \over 2 \pi} \, = \, 0
\end{equation}

 We are going to obtain a graded linear system for the 
determination of the time derivatives $S_{T^{\nu}}$, 
$A_{T^{\nu}}$ in the graded form. Using the facts

$$\Psi_{(0)} (\theta, X, T, 0) \, = \, \Phi \left(
{S (X, T) \over \epsilon} + \theta, k, A, n \right) $$
$$ \Lambda_{(0)} (\theta, X, T, 0) \, = \,
{\omega^{\nu} (k, A, n) \over \epsilon} \,\, \Phi_{\theta} 
\left( {S (X, T) \over \epsilon} + \theta, k, A, n \right) $$
where $\omega^{\nu} (k, A, n)$ is the frequency corresponding 
to the flow $f^{\nu}$ on the space of the one-phase solutions of 
KdV we get from equation (\ref{Equation1}) 
$S_{T^{\nu}} = \omega^{\nu} (k, A, n)$ at $T^{\nu} = 0$
in the main approximation.

 Using also relations (\ref{PsiLambda}) we can write actually

$$S_{T^{\nu}} \, = \, \omega^{\nu} (k, A, n) \, + \,
{\cal O} (\epsilon^{2})$$
at $T^{\nu} = 0$ for the derivative $S_{T^{\nu}}$.

 After the differentiation of the second relation w.r.t. $T^{\nu}$
at $T^{\nu} = 0$ we get the following relation

$$A_{T^{\nu}} \int_{0}^{2\pi} \xi_{1} \, \Phi_{A} \,
{d \theta \over 2 \pi} \, = \, \int_{0}^{2\pi}
\left( {1 \over \epsilon} \, S_{T^{\nu}} \, \xi_{1\theta} \, + \,
k_{T^{\nu}} \, \xi_{1 k} \, + \, A_{T^{\nu}} \,  \xi_{1 A} \, + \,
n_{T^{\nu}} \, \xi_{1 n} \right) \sum_{l \geq 1} 
\epsilon^{l} \, \Phi_{(l)} \,\, {d \theta \over 2 \pi} \, + $$
\begin{equation}
\label{Equation2}
+ \, \int_{0}^{2\pi} \xi_{1} \left( \sum_{l \geq 0}
\epsilon^{2l+1} \, \Lambda_{(2l+1)} |_{(T^{\nu}=0)} \, - \,
k_{T^{\nu}} \, \Phi_{k} \, - \, n_{T^{\nu}} \, \Phi_{n} \right)
\, {d \theta \over 2 \pi} 
\end{equation}
at $T^{\nu} = 0$.

 The function 
$\int_{0}^{2\pi} \xi_{1} \, \Phi_{A} \, d \theta / 2 \pi$
is a strictly positive function on the space of parameters
$(k, A, n)$. Using this fact it is not difficult to see then
that the form of the linear system 
(\ref{Equation1})-(\ref{Equation2}) defines the unique 
representation of the derivatives $S_{T^{\nu}}$, $A_{T^{\nu}}$
at $T^{\nu} = 0$ in the graded form being uniquely resolvable
at every step of the determination of $S_{T^{\nu}}^{[s]}$ and
$A_{T^{\nu}}^{[s]}$. Using also relations (\ref{PsiLambda})
it's not difficult to prove that we obtain the "purely
dispersive" system (\ref{kAnTnu}) in this situation.

 For $0 < T^{\nu} < \delta$ system 
(\ref{Equation1})-(\ref{Equation2}) is still resolvable with
respect to the derivatives $S_{T^{\nu}}$, $A_{T^{\nu}}$, such
that we have

\begin{eqnarray}
\label{Schange1}
&&
S_{T^{\nu}} \,\, = \,\, 
\sum_{l\geq0} \epsilon^{2l} \, \omega_{(2l)}^{\nu} 
(T^{\nu}, k_{0}, A_{0}, n_{0}, k_{0X}, A_{0X}, n_{0X}, \dots )
\nonumber\\
&&
k_{T^{\nu}} \,\, = \,\, \left(
\sum_{l\geq0} \epsilon^{2l} \, \omega_{(2l)}^{\nu} 
(T^{\nu}, k_{0}, A_{0}, n_{0}, k_{0X}, A_{0X}, n_{0X}, \dots )
\right)_{X}
\nonumber\\
&&
\label{kAnchange1}
A_{T^{\nu}} \,\, = \,\,
\sum_{l\geq0} \epsilon^{2l} \, \alpha_{(2l+1)}^{\nu} 
(T^{\nu}, k_{0}, A_{0}, n_{0}, k_{0X}, A_{0X}, n_{0X}, \dots )
\nonumber\\
&&
n_{T^{\nu}} \,\, = \,\,
\sum_{l\geq0} \epsilon^{2l} \, \eta_{(2l+1)}^{\nu} 
(T^{\nu}, k_{0}, A_{0}, n_{0}, k_{0X}, A_{0X}, n_{0X}, \dots ).
\nonumber
\end{eqnarray}
where all the functions $\omega_{(2l)}^{\nu}$, 
$\alpha_{(2l+1)}^{\nu}$, $\eta_{(2l+1)}^{\nu}$ become dependent
on $T^{\nu}$ and do not coincide with the functions from 
(\ref{kAnTnu}) since all the functions 
$\Psi_{(l)}(\theta, X, T, T^{\nu})$ become different from
$\Phi_{(l)}(\theta, X, T)$. However, if we represent the solutions 
of this system, say, in the formal graded form we will be able to
"re-expand" formal solution (\ref{nuseries}) according to change
of parameters (\ref{Tnuparchange}).

 Thus we can write now the new formal graded expansion for
series (\ref{nuseries}) 

\begin{equation}
\label{reexpansion}
\varphi \left( \theta, X, T, T^{\nu} \right) \,\, = \,\, 
\sum_{l\geq0} \epsilon^{l} \, {\tilde{\Psi}}_{(l)} \left(
{S (X, T) \over \epsilon} \, + \, \theta , X, T, T^{\nu} \right)
\end{equation}
according to change of parameters of expansion 
(\ref{Tnuparchange}) at every $T^{\nu}$. System 
(\ref{gradsystems}) can also be easily rewritten for the 
functions ${\tilde{\Psi}}_{(l)} ( \theta , X, T, T^{\nu} )$

$${d \over d T^{\nu}} {\tilde{\Psi}}_{(l)} (\theta , X, T, T^{\nu})
\,\, = \,\, {\tilde{\Lambda}}_{(l)} \left( T^{\nu},
{\tilde{\bm{\Psi}}}, {\tilde{\bm{\Psi}}}_{\theta},
{\tilde{\bm{\Psi}}}_{X}, \dots , 
k, A, n, k_{X}, A_{X}, n_{X}, \dots \right)$$
using system (\ref{kAnchange1}) in this situation. As a
result we will get asymptotic series (\ref{reexpansion}) satisfying
all the conditions (I$^{\prime}$)-(III$^{\prime}$).

 Finally we get that the asymptotic series (\ref{reexpansion}) 
gives a formal solution of (\ref{ekdv}) satisfying all the
conditions (I$^{\prime}$)-(III$^{\prime}$) at $0 < T^{\nu} < \delta$.
As we saw above, solutions (\ref{reexpansion}) should coincide
in this case with the formal graded solution (\ref{phiepsexp})
so we get the invariance of the family (\ref{phiepsexp}) under
the higher KdV flows. The evolution of parameters $(k, A, n)$
is ruled then by system (\ref{kAnTnu}) for all $T^{\nu} > 0$ and 
the commutativity of (\ref{kAnTnu}) with (\ref{epskan}) follows
directly from the commutativity of (\ref{epshigherKdV}) and
(\ref{ekdv}). 

 At last, let us note now that for our conditions 
$k \rightarrow 0$, $X \rightarrow \pm \infty$ we have also
$S_{T^{\nu}}(X) \rightarrow 0$, $X \rightarrow \pm \infty$,
which gives also the conservation of condition (\ref{Skcond})
in our situation.

{\hfill Theorem is proved.}

\vspace{0.5cm}

 It's not difficult to see also that system (\ref{kAnTnu})
coincides with the deformed Whitham system for the higher KdV flow
(\ref{epshigherKdV}) defined by the same normalization conditions
(I$^{\prime}$)-(III$^{\prime}$).

\vspace{0.5cm}

 Let us discuss now the Hamiltonian properties of system 
(\ref{epskan}) following from the Hamiltonian properties of
the KdV equation (\ref{ekdv}). According to the general ideology
of the deformation of systems of Hydrodynamic Type we will
assume the existence of Hamiltonian structures for the deformed
Whitham system given by the deformations of the Hamiltonian
structures of Hydrodynamic Type, i.e. the Hamiltonian structures
having the form

$$\{U^{\nu}(X), U^{\mu}(Y)\} \,\, = \,\,
\{U^{\nu}(X), U^{\mu}(Y)\}_{0} \,\, + $$
\begin{equation}
\label{DeformBracket}
+ \,\, \sum_{k\geq2}
\epsilon^{k-1} \, \sum_{s=0}^{k}
B^{\nu\mu}_{(k)s}({\bf U}, {\bf U}_{X}, \dots, {\bf U}_{(k-s)X})
\,\, \delta^{(s)} (X - Y)
\end{equation}
where all $B^{\nu\mu}_{(k)s}$ are polynomial w.r.t.
derivatives ${\bf U}_{X}$, $\dots$, ${\bf U}_{(k-s)X}$ and
have degree $(k-s)$.

 We call deformations of Hamiltonian structure of
form (\ref{DeformBracket}) the deformations of Dubrovin-Zhang type. 
Bracket (\ref{DeformBracket}) gives a deformation
of the local homogeneous bracket of Hydrodynamic Type 
(Dubrovin - Novikov bracket) which according to the definition 
has the following form

\begin{equation}
\label{DNbr}
\{U^{\nu}(X), U^{\mu}(Y)\} \,\, = \,\, g^{\nu\mu}({\bf U}) \,
\delta^{\prime}(X-Y) \, + \, b ^{\nu\mu}_{\lambda}({\bf U}) \,
U^{\lambda}_{X} \, \delta (X-Y)
\end{equation}

 The corresponding Hamiltonian operator ${\hat J}^{\nu\mu}$
can be written as

$${\hat J}^{\nu\mu} \, = \,  g^{\nu\mu}({\bf U})
{\partial \over \partial x} + b^{\nu\mu}_{\lambda} ({\bf U}) \,
U^{\lambda}_{X} $$
and is homogeneous w.r.t. transformation $X \rightarrow aX$.

 Every functional $H$ of Hydrodynamic Type, i.e. the functional
having the form
 
$$H \, = \, \int_{-\infty}^{+\infty} h({\bf U}) \, dX $$
generates the system of Hydrodynamic Type 

\begin{equation}
\label{HTsyst}
U^{\nu}_{T} \, = \, V^{\nu}_{\mu} ({\bf U}) \, U^{\mu}_{X}
\,\,\,\,\, , \,\,\,\,\, \nu, \, \mu \, = \, 1, \dots, N
\end{equation}
where $V^{\nu}_{\mu} (U)$ is some $N \times N$ matrix
depending on the variables $U^{1}, \dots, U^{N}$
according to the formula   

\begin{equation}
\label{HThamsyst}
U^{\nu}_{T} \, = \, {\hat J}^{\nu\mu} \,
{\delta H \over \delta U^{\mu}(X)} \, = \, g^{\nu\mu}({\bf U})
{\partial \over \partial x} {\partial h \over \partial U^{\mu}}
\, + \, b^{\nu\mu}_{\lambda} ({\bf U}) \,
{\partial h \over \partial U^{\mu}} \, U^{\lambda}_{X}
\end{equation}
 
 The DN-bracket (\ref{DNbr}) is called non-degenerate if
$det \, ||g^{\nu\mu}({\bf U})|| \, \neq \, 0$.

 As was shown by B.A. Dubrovin and S.P. Novikov the theory
of DN-brackets is closely connected with Riemannian
geometry (\cite{dn1,dn2,dn3}). In fact, it follows from the
skew-symmetry of (\ref{DNbr}) that the coefficients
$g^{\nu\mu}({\bf U})$ give in the non-degenerate case the   
contravariant pseudo-Riemannian metric on the manifold
${\cal M}^{N}$ with coordinates $(U^{1}, \dots, U^{N})$
while the functions
$\Gamma^{\nu}_{\mu\lambda}({\bf U}) \, = \,
- g_{\mu\alpha}({\bf U}) \, b^{\alpha\nu}_{\lambda}({\bf U})$
(where $g_{\nu\mu}({\bf U})$ is the corresponding metric
with lower indices) give the connection coefficients
compatible with metric $g_{\nu\mu}({\bf U})$. The validity
of Jacobi identity requires then that $g_{\nu\mu}({\bf U})$
is actually a flat metric on the manifold ${\cal M}^{N}$
and the functions $\Gamma^{\nu}_{\mu\lambda}({\bf U})$
give a symmetric (Levi-Civita) connection on ${\cal M}^{N}$   
(\cite{dn1,dn2,dn3}).

 In the flat coordinates $n^{1}({\bf U}), \dots, n^{N}({\bf U})$
the non-degenerate DN-bracket can be written in constant
form:

$$\{n^{\nu}(X), n^{\mu}(Y)\} \, = \, e^{\nu} \,
\delta^{\nu\mu} \, \delta^{\prime} (X-Y) $$
where $e^{\nu} \, = \, \pm 1$.

 The functionals

$$N^{\nu} \, = \, \int_{-\infty}^{+\infty}
n^{\nu}(X) \, dX $$
are the annihilators of the bracket (\ref{DNbr}) and the
functional

$$P \, = \, {1 \over 2} \int_{-\infty}^{+\infty}
\sum_{\nu = 1}^{N} e^{\nu} \, \left(n^{\nu}(X)\right)^{2}
\, dX $$
is the momentum functional generating the system
$U^{\nu}_{T} \, = \, U^{\nu}_{X}$ according to
(\ref{HThamsyst}).

 The Symplectic Structure corresponding to non-degenerate     
DN-bracket has the weakly nonlocal form and can be written as

$$\Omega_{\nu\mu}(X,Y) \, = \, e^{\nu} \, \delta_{\nu\mu} \,
\nu (X-Y) $$
in coordinates $n^{\nu}$ or, more generally,

$$\Omega_{\nu\mu}(X,Y) \, = \, \sum_{\lambda = 1}^{N}
e^{\lambda} \,
{\partial n^{\lambda} \over \partial U^{\nu}}(X) \,
\nu (X-Y) \,
{\partial n^{\lambda} \over \partial U^{\mu}}(Y) $$
in arbitrary coordinates $U^{\nu}$.

 Let us mention also that the degenerate brackets (\ref{DNbr})
are more complicated but also have a nice differential  
geometric structure (\cite{grinberg}).

 Brackets (\ref{DNbr}) are closely connected with the
integration theory of systems of Hydrodynamic Type
(\ref{HTsyst}). Namely, according to conjecture of
S.P. Novikov, all the diagonalizable systems (\ref{HTsyst})
which are Hamiltonian with respect to DN-brackets (\ref{DNbr})
(with Hamiltonian function of Hydrodynamic Type) are
completely integrable. This conjecture was proved by
S.P. Tsarev (\cite{tsarev}) who proposed a general procedure
("generalized Hodograph method") of integration of
Hamiltonian diagonalizable systems (\ref{HTsyst}).

 In fact Tsarev's "generalized Hodograph method" permits
to integrate the wider class of diagonalizable systems
(\ref{HTsyst}) (semi-Hamiltonian systems, \cite{tsarev}) 
which appeared to be Hamiltonian in more general
(weakly nonlocal) Hamiltonian formalism.

 The corresponding Poisson brackets (Mokhov - Ferapontov
bracket and Ferapontov bracket) are the weakly nonlocal
generalizations of DN-bracket (\ref{DNbr}) and are connected
with geometry of submanifolds in pseudo-Euclidean spaces.
Let us describe here the corresponding structures.

\vspace{0.5cm}

 The Mokhov - Ferapontov bracket
(MF-bracket) has the form (\cite{mohfer1})

\begin{equation}
\label{MFbr}
\{U^{\nu}(X), U^{\mu}(Y)\} \, = \, g^{\nu\mu}({\bf U}) \,
\delta^{\prime} (X-Y) \, + \, b^{\nu\mu}_{\lambda} ({\bf U}) \,
U^{\lambda}_{X} \, \delta (X-Y) \, +
\, c \, U^{\nu}_{X} \, \nu (X-Y) \, U^{\mu}_{Y}   
\end{equation}

 As was proved in \cite{mohfer1} the expression (\ref{MFbr})
with $det \, ||g^{\nu\mu}({\bf U})|| \, \neq \, 0$ gives the
Poisson bracket on the space $U^{\nu}(X)$ if and only if:

1) The tensor $g^{\nu\mu}({\bf U})$ represents the
pseudo-Riemannian contravariant metric of constant curvature
$c$ on the manifold ${\cal M}^{N}$, i.e.

$$R^{\nu\mu}_{\lambda\eta}({\bf U}) \, = \, c \, \left(
\delta^{\nu}_{\lambda} \, \delta^{\mu}_{\eta} \, - \,
\delta^{\mu}_{\lambda} \, \delta^{\nu}_{\eta} \right) $$

2) The functions
$\Gamma^{\nu}_{\mu\lambda}({\bf U}) \, = \,
- g_{\mu\alpha}({\bf U}) \, b^{\alpha\nu}_{\lambda}({\bf U})$
represent the Levi-Civita connection of metric
$g_{\nu\mu}({\bf U})$.

\vspace{0.5cm}

 The Ferapontov bracket (F-bracket) is more general
weakly nonlocal generalization of DN-bracket having the
form (\cite{fer1,fer2,fer3,fer4}):

$$\{U^{\nu}(X), U^{\mu}(Y)\} \, = \, g^{\nu\mu}({\bf U}) \,
\delta^{\prime} (X-Y) \, + \, b^{\nu\mu}_{\lambda}({\bf U}) \,
U^{\lambda}_{X} \, \delta (X-Y) \, + $$
\begin{equation}
\label{Fbr}
+ \, \sum_{k=1}^{g} e_{k} \, w^{\nu}_{(k)\lambda}({\bf U}) \,
U^{\lambda}_{X} \, \nu (X-Y) \,
w^{\mu}_{(k)\delta}({\bf U}) \, U^{\delta}_{Y}
\end{equation}
$e_{k} = \pm 1$, $\nu, \mu = 1, \dots, N$.

 The expression (\ref{Fbr})
(with $det \, ||g^{\nu\mu}({\bf U})|| \, \neq \, 0$) gives the
Poisson bracket on the space $U^{\nu}(X)$ if and only if
(\cite{fer1,fer4}):

1) Tensor $g^{\nu\mu}({\bf U})$ represents the metric
of the submanifold ${\cal M}^{N} \subset {\mathbb E}^{N+g}$
with flat normal connection in the pseudo-Euclidean space
${\mathbb E}^{N+g}$ of dimension ${N+g}$;

2) The functions
$\Gamma^{\nu}_{\mu\lambda}({\bf U}) \, = \,
- g_{\mu\alpha}({\bf U}) \, b^{\alpha\nu}_{\lambda}({\bf U})$
represent the Levi-Civita connection of metric
$g_{\nu\mu}({\bf U})$;

3) The set of affinors $\{w^{\nu}_{(k)\lambda}({\bf U})\}$
represents the full set of Weingarten operators corresponding
to $g$ linearly independent parallel vector fields in the
normal bundle, such that:

$$g_{\nu\tau}({\bf U}) \, w^{\tau}_{(k)\mu}({\bf U}) \, = \,
g_{\mu\tau}({\bf U}) \, w^{\tau}_{(k)\nu}({\bf U})
\,\,\,\,\, , \,\,\,\,\,
\nabla_{\nu} \, w^{\mu}_{(k)\lambda}({\bf U}) \, = \,
\nabla_{\lambda} \, w^{\mu}_{(k)\nu}({\bf U}) $$

$$R^{\nu\mu}_{\lambda\eta}({\bf U}) \, = \,
\sum_{k=1}^{g} e_{k} \, \left(
w^{\nu}_{(k)\lambda}({\bf U}) \,
w^{\mu}_{(k)\eta}({\bf U}) \, - \,
w^{\mu}_{(k)\lambda}({\bf U}) \,
w^{\nu}_{(k)\eta}({\bf U}) \right) $$

 Besides that the set of affinors $w_{(k)}$ is commutative
$[w_{(k)},w_{(k^{\prime})}]  = 0$.

 As was shown in \cite{fer2} the expression (\ref{Fbr}) can  
be considered as the Dirac reduction of the Dubrovin-Novikov
bracket connected with metric in ${\mathbb E}^{N+g}$ to the
manifold ${\cal M}^{N}$ with flat normal connection. Let
us note also that MF-bracket can be considered as a case  
of the F-bracket when ${\cal M}^{N}$ is a (pseudo)-sphere
${\cal S}^{N} \subset {\mathbb E}^{N+1}$ in a pseudo-Euclidean
space.

\vspace{0.5cm}

 The Symplectic Structures $\Omega_{\nu\mu}(X,Y)$ for
both (non-degenerate) MF-bracket and F-bracket have also
the weakly nonlocal form (\cite{PhysD})
and can be written in general coordinates $U^{\nu}$ as

$$\Omega_{\nu\mu}(X,Y) \, = \, \sum_{s=1}^{N+g}
\epsilon_{s} \,
{\partial n^{s} \over \partial U^{\nu}}(X) \,
\nu (X-Y) \,
{\partial n^{s} \over \partial U^{\mu}}(Y) $$
where $\epsilon_{s} \, = \, \pm 1$ and the metric $G_{IJ}$
in the space ${\mathbb E}^{N+g}$ has the form
$G_{IJ} \, = \, diag (\epsilon_{1}, \dots, \epsilon_{N+g})$.
The functions $n^{1}({\bf U}), \dots, n^{N+g}({\bf U})$
are the "Canonical forms" on the manifold ${\cal M}^{N}$
and play the role of densities and annihilators of bracket
(\ref{Fbr}) and "Canonical Hamiltonian functions"
(see \cite{PhysD}) depending on the definition of phase
space. In fact, the functions $n^{s}({\bf U})$ are the 
restrictions of flat coordinates of metric $G_{IJ}$    
giving the DN-bracket in ${\mathbb E}^{N+g}$ on manifold
${\cal M}^{N}$. The mapping
${\cal M}^{N} \rightarrow {\mathbb E}^{N+g}$:

$$\left( U^{1}, \dots, U^{N}\right) \,\,\,\,\,\,\,
\rightarrow \,\,\,\,\,\,\,
\left( n^{1}({\bf U}), \dots,  n^{N+g}({\bf U}) \right) $$
gives locally the embedding of ${\cal M}^{N}$ in
${\mathbb E}^{N+g}$ as a submanifold with flat normal connection.

\vspace{0.5cm}

 All the brackets (\ref{DNbr}), (\ref{MFbr}), (\ref{Fbr})
are connected with Tsarev method of integration of systems
(\ref{HTsyst}). Namely, any diagonalizable system (\ref{HTsyst})
Hamiltonian w.r.t. the (non-degenerate) bracket
(\ref{DNbr}), (\ref{MFbr}) or (\ref{Fbr}) can be integrated
by "generalized Hodograph method".

 We will not describe here Tsarev method in details. However,
let us point out that "generalized Hodograph method" and the 
HT Hamiltonian Structures were very useful for Whitham's
systems obtained by the averaging of integrable PDE's   
(\cite{whith1,ffm,dn1,krichev1,krichev2,dn2,dn3}).

 The Hamiltonian approach to the Whitham method was started
by B.A. Dubrovin and S.P. Novikov
in \cite{dn1} (see also \cite{dn2,dn3}) where the procedure  
of "averaging" of local field-theoretical Poisson bracket
was proposed. The Dubrovin - Novikov procedure gives the
DN-bracket for the Whitham system (\ref{HTsyst}) in case
when the initial system is Hamiltonian w.r.t. a local Poisson 
bracket

$$\{\varphi^{i}(x), \varphi^{j}(y)\} \, = \, \sum_{k \geq o}   
B_{(k)}^{ij} (\bm{\varphi}, \bm{\varphi}_{x}, \dots) \,   
\delta^{(k)}(x-y)$$
with local Hamiltonian functional

$$H \, = \, \int_{-\infty}^{+\infty}
h (\bm{\varphi}, \bm{\varphi}_{x}, \dots) \, dx $$

 The method of Dubrovin and Novikov is based on the presence
of $N$ (equal to the number of parameters $U^{\nu}$ of the
family of $m$-phase solutions) local integrals
  
\begin{equation}
\label{integ}
I^{\nu} = \int {\cal P}^{\nu}
(\bm{\varphi},\bm{\varphi}_{x},\dots) dx
\end{equation}
commuting with the Hamiltonian function and with each other 

\begin{equation}
\label{invv}
\{I^{\nu} , H\} = 0 \,\,\,\,\, , \,\,\,\,\,
\{I^{\nu} , I^{\mu}\} = 0
\end{equation}
and can be formulated in the following form:

  We calculate the pairwise Poisson brackets of the
densities ${\cal P}^{\nu}$ in the form

$$\{{\cal P}^{\nu}(x), {\cal P}^{\mu}(y)\} =
\sum_{k\geq 0} A^{\nu\mu}_{k}
(\bm{\varphi},\bm{\varphi}_{x},\dots)
\delta^{(k)}(x-y) $$
where
$$A^{\nu\mu}_{0}(\bm{\varphi},\bm{\varphi}_{x},\dots) \equiv
\partial_{x} Q^{\nu\mu}(\bm{\varphi},\bm{\varphi}_{x},\dots) $$ 
according to (\ref{invv}). Then the Dubrovin-Novikov bracket
on the space of functions $U(X)$ can be written in the form

\begin{equation}
\label{dubrnovb}
\{U^{\nu}(X), U^{\mu}(Y)\} =
\langle A^{\nu\mu}_{1}\rangle (U) \,\, \delta^{\prime}(X-Y) +
{\partial \langle Q^{\nu\mu} \rangle \over
\partial U^{\gamma}} U^{\gamma}_{X} \,\, \delta (X-Y)
\end{equation}  
where $\langle \dots \rangle$ means the averaging on the
family of $m$-phase solutions given by the
formula:

$$\langle F \rangle \, = \, \lim_{c \rightarrow \infty}
{1 \over 2c} \int_{-c}^{c} F (\varphi,\varphi_{x},\dots)
dx \, = \, {1 \over (2\pi)^{m}} 
\int_{0}^{2\pi}\!\!\dots\int_{0}^{2\pi}
F (\Phi, k^{\alpha}(U) \Phi_{\theta^{\alpha}}, \dots)
d^{m}\theta $$
and we choose the parameters $U^{\nu}$ such that they coincide 
with the values of $I^{\nu}$ on the corresponding solutions   

$$U^{\nu} = \langle P^{\nu}(x) \rangle $$

 This procedure was generalized in \cite{malnloc2}
for the weakly nonlocal Hamiltonian structures. In this case
the procedure of construction of general F-bracket (or MF-bracket) 
for the Whitham system from the weakly non-local Poison bracket 
for initial system was proposed.\footnote{The final proof of the
Jacobi identity for the bracket given by the Dubrovin - Novikov
procedure was given in \cite{izvestia}.}
In \cite{sympform} the procedure of averaging of the
weakly-nonlocal Symplectic structures was also suggested.

 Here we are going to consider the construction of 
bracket (\ref{DeformBracket}) for the deformed Whitham system
(\ref{epskan}). Being considered for the KdV case the
procedure will have in fact general character and can be
considered as a generalization of the Dubrovin - Novikov 
procedure for the deformed Whitham system in general case.

 We are going to use the Dirac restriction of a Poisson
bracket on a submanifold to establish the procedure of the
construction of a Poisson bracket of form (\ref{DeformBracket})
for the deformed Whitham system (\ref{epskan}) which we call
the "averaging" of a Poisson bracket for the deformed Whitham 
systems. Let us first introduce the Poisson brackets

\begin{equation}
\label{ExtGZF}
\{ \varphi (\theta, X), \varphi (\theta^{\prime}, Y) \} 
\,\, = \,\,
\epsilon \, \delta (\theta - \theta^{\prime}) \,
\delta^{\prime} (x-y) 
\end{equation}
and
  
\begin{equation}
\label{ExtMagri}
\{ \varphi (\theta, X), \varphi (\theta^{\prime}, Y) \}
\,\, = 
\end{equation}
$$ = \,\, \epsilon^{3} \, \delta (\theta - \theta^{\prime}) \,
\delta^{\prime\prime\prime} (X - Y) \, + \, \epsilon \,
{2 \over 3} \, \varphi (X) \, \delta (\theta - \theta^{\prime}) 
\, \delta^{\prime} (X - Y) \, + \, \epsilon \,
{1 \over 3} \, \varphi_{X} \, 
\delta (\theta - \theta^{\prime}) \, \delta (X - Y) $$
which correspond to the Gardner - Zakharov - Faddeev and the
Magri brackets on the extended phase space $\varphi (\theta, X)$
periodic in $\theta$ with the period $2\pi$. Easy to see that both
the expressions (\ref{ExtGZF}) and (\ref{ExtMagri}) give Poisson
brackets on the extended functional space.

 We have to consider now the "subspace" in the extended functional 
space corresponding to the full family (\ref{phiepsexp}) 
parametrized by three functional parameters $(S(X), A(X), n(X))$.
\footnote{Let us note that all the constructions are considered 
here just on the level of the formal asymptotic series.} We will
call now the Dirac restriction of bracket (\ref{ExtGZF}) or
(\ref{ExtMagri}) on the submanifold corresponding to the full
family of solutions (\ref{phiepsexp}) the averaging of the
Gardner - Zakharov - Faddeev bracket or the Magri bracket
giving a Poisson bracket for the deformed Whitham system
(\ref{epskan}).

 Let us remind that the Dirac restriction of a Poisson bracket
on a submanifold ${\cal N}^{k} \subset {\cal M}^{n}$ is connected
with the special choice of coordinates in the vicinity of the
submanifold ${\cal N}^{k}$ which are divided to the "coordinates
on the submanifold" $(U^{1}, \dots, U^{k})$ and the constraints
$(g^{1}, \dots, g^{n-k})$ which define the submanifold 
${\cal N}^{k}$. It is assumed that the submanifold ${\cal N}^{k}$
is given by the conditions

$$g^{i} ({\bf x}) \, = \, 0 \,\,\,\,\, , \,\,\,\,\,
i = 1, \dots, n - k $$
while the $k$ functions $U^{1} ({\bf x})$, $ \dots $,
$U^{k} ({\bf x})$ on ${\cal M}^{n}$ play the role of coordinate
system on ${\cal N}^{k}$ after the restriction on this
submanifold.

 If the Hamiltonian flows generated by the functions 
$U^{j} ({\bf x})$ leave the submanifold ${\cal N}^{k}$ invariant,
i.e. we have

$$\{ U^{j} ({\bf x}) , g^{i} ({\bf x}) \} \,\, = \,\, 0
\,\,\,\,\, {\rm for}  \,\,\,\,\, {\bf g} ({\bf x}) \, = \, 0 $$
then the pairwise Poisson brackets of functions $U^{j} ({\bf x})$
give a Poisson tensor after the restriction on ${\cal N}^{k}$
with coordinates $(U^{1}, \dots, U^{k})$ which is called the
Dirac restriction of the Poisson bracket $\{ \dots , \dots \}$
defined on ${\cal M}^{n}$ on the submanifold 
${\cal N}^{k} \subset {\cal M}^{n}$.

 In general, according to Dirac procedure, if we have some 
constraints $g^{i} ({\bf x})$ which define a submanifold
${\cal N}^{k}$ and some functions $U^{j} ({\bf x})$ giving
a coordinate system on ${\cal N}^{k}$ we have to find $k$
linear combinations 
$\beta^{j}_{s}({\bf U}) \, g^{s} ({\bf x})$ at every point of
${\cal N}^{k}$ such that we have for the functions

$${\tilde{U}}^{j} ({\bf x}) \,\, = \,\, U^{j} ({\bf x}) \, + \,
\beta^{j}_{s}({\bf U}) \, g^{s} ({\bf x}) \,\,\,\,\, , \,\,\,\,\,
j = 1, \dots, k $$
the relations

$$\{ {\tilde{U}}^{j} ({\bf x}) , g^{i} ({\bf x}) \} \,\, = \,\, 0
\,\,\,\,\, {\rm at}  \,\,\,\,\, {\bf g} ({\bf x}) \, = \, 0 $$

 The functions ${\tilde{U}}^{j} ({\bf x})$ have the same values
as the functions $U^{j} ({\bf x})$ at the points of ${\cal N}^{k}$
and we can then define the Dirac bracket $\{ \dots , \dots \}_{D}$
on ${\cal N}^{k}$ by the formula

$$\{ U^{i} , U^{j} \}_{D} \,\, = \,\, 
\{ {\tilde{U}}^{i} ({\bf x}) , {\tilde{U}}^{j} ({\bf x}) \}
|_{{\cal N}^{k}} ({\bf U}) $$

 The functions $\beta^{j}_{s}({\bf U})$ are defined from the
linear system

$$\{ g^{i} ({\bf x}) , g^{s} ({\bf x}) \}|_{{\cal N}^{k}} \,
\beta^{j}_{s}({\bf U}) \,\, + \,\,
\{ g^{i} ({\bf x}) , U^{j} ({\bf x}) \}|_{{\cal N}^{k}} 
\,\,\, = \,\,\, 0 \,\,\,\,\, , \,\,\,\,\, i = 1, \dots, n-k $$
and we can also write

$$\{ U^{i} , U^{j} \}_{D} \,\, = \,\,
\{ U^{i} ({\bf x}) , U^{j} ({\bf x}) \}|_{{\cal N}^{k}}
\,\, - \,\, \beta^{i}_{s}({\bf U}) \, 
\{ g^{s} ({\bf x}) , g^{q} ({\bf x}) \}|_{{\cal N}^{k}} \,
\beta^{j}_{q}({\bf U}) $$
for the Dirac bracket on ${\cal N}^{k}$.

 Let us describe now the Dirac procedure in our situation.

 First, let us introduce new coordinates on the submanifold
${\cal K}$ corresponding to solutions (\ref{phiepsexp})
based on the conservation laws of the KdV equation (\ref{ekdv}).

 Let us choose three integrals of the KdV equation such that
their values on the family of one-phase solutions of KdV are
functionally independent. In our case it is most convenient to
take the integrals

$$I_{0} \, = \, \int_{-\infty}^{+\infty} \varphi \,\, dx 
\,\,\, , \,\,\,
I_{1} \, = \, \int_{-\infty}^{+\infty} {\varphi^{2} \over 2} 
\,\, dx \,\,\, , \,\,\,
I_{2} \, = \, \int_{-\infty}^{+\infty} \left(
{\varphi^{2} \over 6} - {\varphi_{x}^{2} \over 2} \right) 
\, dx $$

 The integrals transform naturally to the integrals of the
KdV equation (\ref{ekdv}) on the extended phase space

$$I_{0} \, = \, \int_{-\infty}^{+\infty} \int_{0}^{2\pi}
\varphi \,\, d X \, {d \theta \over 2\pi}
\,\,\, , \,\,\,
I_{1} \, = \, \int_{-\infty}^{+\infty} \int_{0}^{2\pi}
{\varphi^{2} \over 2}
\,\, d X \, {d \theta \over 2\pi} $$
$$ I_{2} \, = \, \int_{-\infty}^{+\infty} \int_{0}^{2\pi} 
\left( {\varphi^{2} \over 6} - {\epsilon^{2} \over 2}
\varphi_{X}^{2} \right) \, d X \, {d \theta \over 2\pi} $$

 Let us introduce now the functionals

$$U^{\nu} (X) \,\, = \,\, \int_{0}^{2\pi} {\cal P}^{\nu}
(\varphi, \epsilon \varphi_{X}, \dots ) \,\, 
{d \theta \over 2\pi} \,\,\,\,\, , \,\,\,\,\,
\nu = 0, 1, \dots $$
i.e.

$$U^{0} (X) \,\, = \,\, \int_{0}^{2\pi}
\varphi (\theta, X) \,\, {d \theta \over 2\pi}
\,\,\,\,\,\,\,\ , \,\,\,\,\,\,\,\, 
U^{1} (X) \,\, = \,\, {1 \over 2} \int_{0}^{2\pi}
\varphi^{2} (\theta, X) \,\, {d \theta \over 2\pi} $$
\begin{equation}
\label{U012}
U^{2} (X) \,\, = \,\, \int_{0}^{2\pi}
\left( {\varphi^{2} \over 6} (\theta, X) - {\epsilon^{2} \over 2}
\varphi_{X}^{2} (\theta, X) \right) \,\, {d \theta \over 2\pi} 
\end{equation}
and consider the values of the functionals $U^{\nu} (X)$ on the
submanifold ${\cal K}$.

 It is easy to see that the values of $U^{\nu} (X)$ on ${\cal K}$
are equal in the main approximation to the values of the
functionals $I^{\nu}$ on the one-phase solutions of KdV with
the parameters $(k (X), A (X), n (X))$ and have in general
higher corrections polynomial in derivatives of the functions
$k (X)$, $A (X)$, and $n (X)$. It's not difficult to see also
that the higher corrections to $U^{\nu} (X)$ contain only even
degrees in the expansion w.r.t. the derivatives $k_{X}$, 
$A_{X}$, $n_{X}$, $\dots$ for our choice of the initial phase
of the functions $\Phi (\theta, k, A, n)$ view the statements
of the Lemma 3.1. Thus for the functionals $U^{0} (X)$,
$U^{1} (X)$, $U^{2} (X)$ we can write

$$U^{0} (X) \,\, = \,\, \langle \varphi \rangle_{0} (X) 
\,\, \equiv \,\, n (X) $$
\begin{equation}
\label{UkAnrel}
U^{1} (X) \,\, = \,\, \langle \varphi^{2} \rangle_{0} (X)
\, + \, \sum_{s\geq1} U^{1}_{2s} 
\end{equation}
$$U^{2} (X) \,\, = \,\, \langle
{\varphi^{2} \over 6} - {\varphi_{x}^{2} \over 2} \rangle_{0} (X)
\, + \, \sum_{s\geq1} U^{2}_{2s} $$
where

$$\langle \varphi \rangle_{0} \,\, \equiv \,\, \int_{0}^{2\pi}
\Phi (\theta, k, A, n) \,\, {d \theta \over 2\pi} 
\,\, \equiv \,\, n $$
$$\langle \varphi^{2} \rangle_{0} \,\, \equiv \,\, \int_{0}^{2\pi}
\Phi^{2} (\theta, k, A, n) \,\, {d \theta \over 2\pi} $$
$$\langle {\varphi^{2} \over 6} - {\varphi_{x}^{2} \over 2} 
\rangle_{0} \,\, = \,\, \int_{0}^{2\pi} \left( {1 \over 6}
\Phi^{3} (\theta, k, A, n) \, - \, {k^{2} \over 2} \,
\Phi_{\theta}^{2} (\theta, k, A, n) \right) 
{d \theta \over 2\pi} $$
and the values $U^{1}_{2s}$, $U^{2}_{2s}$ are graded polynomials
in the derivatives of $k$, $A$, and $n$ having degree $2s$.

 Since the values of $I^{0}$, $I^{1}$, $I^{2}$ are functionally
independent on the space of one-phase solutions of KdV we can
write the "inverted series" for the functions $k (X)$, $A (X)$,
$n (X)$. We have to change the gradation rules now such that we
will define the gradation degree with respect to the
$X$-derivatives of the parameters $(U^{0}, U^{1}, U^{2})$
instead of $(k, A, n)$. So we can write now

$$k (X) \,\, = \,\, k_{0} (U^{0}(X), U^{1}(X), U^{2}(X))
\, + \, \sum_{s\geq1} k_{(2s)}([U^{0}, U^{1}, U^{2}], X) $$
\begin{equation}
\label{kAnUrel}
A (X) \,\, = \,\, A_{0} (U^{0}(X), U^{1}(X), U^{2}(X)) 
\, + \, \sum_{s\geq1} A_{(2s)}([U^{0}, U^{1}, U^{2}], X) 
\end{equation}
$$n (X) \,\, = \,\, U^{0} (X) $$
where $k_{0} (U^{0}, U^{1}, U^{2})$, 
$A_{0} (U^{0}, U^{1}, U^{2})$ are the exact "one-phase" 
expressions for the parameters $k$ and $A$ in terms of
$(U^{0}, U^{1}, U^{2})$ and the functions 
$k_{(2s)}$, $A_{(2s)}$ are graded polynomials in the
derivatives of $(U^{0}, U^{1}, U^{2})$ having degree $2s$.

 Using relations (\ref{kAnUrel}) we can re-expand also solutions
(\ref{phiepsexp}) as graded series with respect to the
$X$-derivatives of the values of the functionals $U^{0} (X)$,
$U^{1} (X)$, $U^{2} (X)$ on ${\cal K}$ at every time, such
that we have

\begin{equation}
\label{Uexpphi}
\phi (\theta, X, T) \,\, = \,\,
\Phi^{\bf U} \left( 
{S (X,T) \over \epsilon} + \theta, U^{0}, U^{1}, U^{2} 
\right) \, + \, \sum_{l\geq1} \epsilon^{l} \, \Phi^{\bf U}_{(l)}
\left( {S (X,T) \over \epsilon} + \theta, X, T \right)
\end{equation}
where all $\Phi^{\bf U}_{(l)}$ are graded polynomials of
$(U^{0}_{X}, U^{1}_{X}, U^{2}_{X}, \dots)$ of degree $l$. 

 The function $\Phi^{\bf U} (\theta, U^{0}, U^{1}, U^{2})$
represents the exact one-phase solution of KdV depending on
parameters $(U^{0}, U^{1}, U^{2})$ and we have by definition

$$\Phi^{\bf U} (\theta, U^{0}, U^{1}, U^{2}) \,\, = \,\,
\Phi (\theta, k ({\bf U}), A ({\bf U}), n ({\bf U})) $$

 According to our approach we will assume that series
(\ref{Uexpphi}) and (\ref{phiepsexp}) are equivalent
representations of formal asymptotic solutions (\ref{phiepsexp})
connected by change of the asymptotic functional parameters
(\ref{UkAnrel})-(\ref{kAnUrel}). Let us note also that due
to the form of relations (\ref{UkAnrel})-(\ref{kAnUrel})
the symmetric properties of the functions 
$\Phi^{\bf U}_{(l)} (\theta, X, T)$ remain the same as for the
terms of series (\ref{phiepsexp}), i.e. we have

$$\Phi^{\bf U}_{(2s)} (-\theta, X, T) \, = \,
\Phi^{\bf U}_{(2s)} (\theta, X, T) \,\,\,\,\, , \,\,\,\,\,
\Phi^{\bf U}_{(2s+1)} (-\theta, X, T) \, = \,
- \, \Phi^{\bf U}_{(2s+1)} (\theta, X, T) 
\,\,\,\,\, , \,\,\,\,\, s \geq 0 $$

 We can assume in the same way that the functions of the 
submanifold ${\cal K}$ are represented now by the asymptotic
series

$$\varphi (\theta, X) \,\, = \,\, \Phi^{\bf U} 
\left( {S (X) \over \epsilon} + \theta, 
U^{0}, U^{1}, U^{2} \right) \, + \, 
\sum_{l\geq1} \epsilon^{l} \, \Phi^{\bf U}_{(l)}
\left( {S (X) \over \epsilon} + \theta, 
[U^{0}, U^{1}, U^{2}], X \right) $$
so the functionals $U^{0}(X)$, $U^{1}(X)$, $U^{2}(X)$ play the
role of coordinates on this submanifold and these are exactly 
the functionals we are going to use for the Dirac procedure.

 Let us introduce now the system of "constraints" which defines 
our submanifold ${\cal K}$ in the functional space. For our
purposes it will be convenient to write the constraints in the
following form:

 Let us denote

$$\psi (\theta, [{\bf U}], X) \,\, = \,\, 
\Phi^{\bf U} (\theta, {\bf U}, X) \, + \, 
\sum_{l\geq1} \epsilon^{l} \, \Phi^{\bf U}_{(l)}
(\theta, [{\bf U}], X) $$
where the notations 
${\bf U (X)} = (U^{0}(X), U^{1}(X), U^{2}(X))$ denote now
the functionals (\ref{U012}) defined on the full functional
space $\{ \varphi (\theta, X) \}$. We introduce now the
constraints $g (\theta, X)$ by the formula

\begin{equation}
\label{gconstr}
g (\theta, X) \,\, = \,\, \varphi (\theta, X) \, - \,
\psi \left( {S [{\bf U}](X) \over \epsilon} + \theta,
[{\bf U}], X \right)
\end{equation}
as the functionals on the space ${\cal M}$. Easy to see that
the relations 

$$g (\theta, X) \,\,\, = \,\,\, 0 $$
define then exactly the "sub-manifold" ${\cal K}$ we consider
here.

 However, the set of constraints (\ref{gconstr}) is certainly
not independent in the ordinary sense. Namely, in the full
analogy with the finite-dimensional case the following relations
take place identically for the "gradients" 
$\delta g (\theta, X) / \delta \varphi (\theta^{\prime}, Y)$
on the "sub-manifold" ${\cal K}$:

\begin{equation}
\label{dependence}
\int_{-\infty}^{+\infty} \int_{0}^{2\pi} 
{\delta U^{\nu} (Z) \over \delta \varphi (\theta, X)} \,\,
{\delta g (\theta, X) \over  \varphi (\theta^{\prime}, Y)} \,\,
{d \theta \over 2\pi} \,\, d X \,\, \equiv \,\, 0
\end{equation}

 Nevertheless, it will be convenient for us not to choose an
independent system of constraints and to keep constraints
(\ref{gconstr}) for our purposes, so we have to remember the
presence of relations (\ref{dependence}) for system 
(\ref{gconstr}).

 For the Dirac restriction of bracket (\ref{ExtGZF}) or
(\ref{ExtMagri}) on the submanifold ${\cal K}$ we have to
modify now the functionals $U^{0}(X)$, $U^{1}(X)$, $U^{2}(X)$
by the linear combinations of constraints $g (\theta, X)$

$${\tilde U}^{\nu} (X) \,\, = \,\, U^{\nu} (X) \, + \,
\int_{-\infty}^{+\infty} \int_{0}^{2\pi}
g (\theta, Y) \,\, \beta ^{\nu} 
\left( {S [{\bf U}](Y) \over \epsilon} + \theta, 
[{\bf U}], Y, X \right) {d \theta \over 2\pi} \, d Y $$
such that the functionals ${\tilde U}^{\nu} (X)$ leave
invariant the submanifold ${\cal K}$ in the corresponding
Hamiltonian structure and then to use the functionals 
${\tilde U}^{\nu} (X)$ for the construction of the Dirac
bracket on ${\cal K}$. The functions
$\beta ^{\nu} (S (Y) / \epsilon + \theta, Y, X)$
should satisfy the relation

\begin{equation}
\label{betasystem}
\int_{-\infty}^{+\infty} \int_{0}^{2\pi}
\{ g (\theta, X) , g (\theta^{\prime}, Z) \} \,
\beta^{\nu} \left( {S (Z) \over \epsilon} + 
\theta^{\prime}, Z, Y \right) {d \theta^{\prime} \over 2\pi}
\, d Z \, + \, \{ g (\theta, X) , U^{\nu} (Y) \}
\, = \, 0
\end{equation}
on ${\cal K}$ and are defined at every "point" of ${\cal K}$
modulo the linear combinations of the functions 
$\delta U^{\mu} (W) / \delta \varphi (\theta, Y)$ view the
original dependence of constraints (\ref{gconstr}).

 The Dirac bracket on the manifold ${\cal K}$ can be defined
by the formula 

\begin{equation} 
\label{DiracBracket}
\{ U^{\nu} (X) , U^{\mu} (Y) \}_{D} \,\, = \,\,
\{ U^{\nu} (X) , U^{\mu} (Y) \}|_{\cal K} \, - 
\end{equation}
$$ - \, \int
\beta^{\nu} \left( {S (X) \over \epsilon} + \theta, Z, X \right)
\{ g (\theta, Z) , g (\theta^{\prime}, W) \}|_{\cal K} \,
\beta^{\mu} \left( {S (Y) \over \epsilon} + \theta^{\prime},
W, Y \right) {d \theta \over 2\pi} {d \theta^{\prime} \over 2\pi}
dZ dW $$
so the procedure gives a unique definition of the bracket
$\{ U^{\nu} (X) , U^{\mu} (Y) \}_{D}$.

 To get a local deformed Poisson bracket on ${\cal K}$ we will 
try to find the functions $\beta^{\nu} (\theta, Y, X)$ in the 
form

\begin{equation}
\label{beta1def}
\beta^{\nu} (\theta, Y, X) \,\, = \,\, \sum_{s\geq1}
\epsilon^{s} \, \beta_{(s)}^{\nu} (\theta, Y, X)
\end{equation}
where the functions $\beta_{(s)}^{\nu} (\theta, Y, X)$ are
represented as the local distributions

\begin{equation}
\label{beta2def}
\beta_{(s)}^{\nu} (\theta, Y, X) \,\, = \,\, \sum_{p=0}^{s}
\beta_{(s),p}^{\nu} (\theta, Y) \, \delta^{(p)} (Y - X)
\end{equation}
having gradation $s$ assuming that the derivatives of the 
delta-function $\delta^{(p)} (Y - X)$ have degree $p$ by
definition. 

 Thus, we assume that all the functions 
$\beta_{(s),p}^{\nu} (\theta, Y)$ on ${\cal K}$ are local
functionals of \linebreak
$(U^{0}(X), U^{1}(X), U^{2}(X), U^{0}_{X}, U^{1}_{X},
U^{2}_{X}, \dots )$ at every $\theta$, polynomial in derivatives
\linebreak $(U^{0}_{X}, U^{1}_{X}, U^{2}_{X}, \dots )$ 
and having degree $s-p$ according to our previous definition. 
This structure of $\beta^{\nu} (\theta, Y, X)$ is obviously 
equivalent to the statement that the functionals 

$$\int_{-\infty}^{+\infty} U^{\nu} (X) \,\, q (X) \,\, d X $$
with a "slow" function of $X$ $q (X)$ can be modified with the
aid of a linear combination of constraints (\ref{gconstr})
with the coefficients

$$B^{\nu}_{[q]} (\theta, Y) \,\, = \,\, \sum_{s\geq1}
\epsilon^{s} \, \sum_{p=0}^{s} \beta_{(s),p}^{\nu} (\theta, Y) 
\, {d^{p} q (X) \over d X^{p}} $$
to leave the submanifold ${\cal K}$ invariant. According to
this scheme the derivatives $d^{s} q / d X^{s}$ of the slow
function $q (X)$ have degree $s$ as well as the derivatives of
the parameters $U^{0}(X)$, $U^{1}(X)$, $U^{2}(X)$.

 Finally, we have to study now system ({\ref{betasystem}) for
the cases of the Gardner - Zakharov - Faddeev bracket and the
Magri bracket to investigate the possibility to find the
functions $\beta^{\nu} (\theta, Y, X)$ in form 
(\ref{beta1def})-(\ref{beta2def}). Let us formulate here the
following Theorem.

\vspace{0.5cm}

{\bf Theorem 4.2.}

{\it Both for the Gardner - Zakharov - Faddeev bracket and
the Magri bracket for KdV the functions 
$\beta^{\nu} (\theta, Y, X)$ can be found in form 
(\ref{beta1def})-(\ref{beta2def}) on the family ${\cal K}$.
Thus, the Dirac restriction of both the brackets on the
family ${\cal K}$ has the local deformed Hydrodynamic form
(\ref{DeformBracket}) which gives two deformed Hydrodynamic
Type brackets for the deformed Whitham system (\ref{epskan}).}

\vspace{0.5cm}

 Proof.

 Let us analyze equations (\ref{betasystem}) for the case of the
Gardner - Zakharov - Faddeev bracket and the Magri bracket. We
have first on the family ${\cal K}$

$$\{g (\theta, X) , g (\theta^{\prime}, Z)\}|_{\cal K} \,\, = \,\, 
\{ \varphi (\theta, X) , \varphi (\theta^{\prime}, Z) \}|_{\cal K}
\,\, - $$
$$- \,\, \int_{-\infty}^{+\infty} d W \,\,
\{ \varphi (\theta, X) , U^{\mu} (W) \}|_{\cal K} \,
{\delta \psi (S [{\bf U}](Z)/\epsilon + \theta^{\prime},
[{\bf U}], Z ) \over \delta U^{\mu} (W)} \,\, - $$
$$- \,\, \int_{-\infty}^{+\infty} d W \,\,
{\delta \psi (S [{\bf U}](X)/\epsilon + \theta ,
[{\bf U}], X ) \over \delta U^{\mu} (W)} \,\,
\{ U^{\mu} (W) , \varphi (\theta^{\prime}, Z) \}|_{\cal K} \,\, + $$
$$+ \!\! \int_{-\infty}^{+\infty} \!\!\!
\int_{-\infty}^{+\infty} \!\! d W d V  
{\delta \psi (S [{\bf U}](X)/\epsilon + \theta ,        
[{\bf U}], X ) \over \delta U^{\mu} (W)} 
\{ U^{\mu} (W) , U^{\gamma} (V) \}|_{\cal K} 
{\delta \psi (S [{\bf U}](Z)/\epsilon + \theta^{\prime}, 
[{\bf U}], Z ) \over \delta U^{\gamma} (V)} $$
(summation over repeated indices).\footnote{Let us note that we
assume the differentiation $\delta / \delta U$ in the sense of the
values of functionals $U^{\nu} (X)$ on the family ${\cal K}$
while we treat $U^{\nu} (X)$ inside the brackets as a 
functional on the whole functional space.}

 In the same way

$$\{g (\theta, X) , U^{\nu} (Y) \}|_{\cal K} \,\, = \,\,
\{ \varphi (\theta, X) , U^{\nu} (Y) \}|_{\cal K} \,\, - $$
$$- \, \int_{-\infty}^{+\infty} d W \,\,
{\delta \psi (S [{\bf U}](X)/\epsilon + \theta ,
[{\bf U}], X ) \over \delta U^{\mu} (W)} \,\,
\{ U^{\mu} (W) , U^{\nu} (Y) \}|_{\cal K} $$

 Let us say now some words about the Poisson bracket
$\{ k (W) , U^{\nu} (Y) \}|_{\cal K}$. As we saw already
the functional

$$I^{\nu} \,\, = \,\, \int_{-\infty}^{+\infty}
U^{\nu} (Y) \,\, d Y $$
leaves invariant the submanifold ${\cal K}$ so the Poisson
bracket $\{ k (W) , I^{\nu} \}|_{\cal K}$ should give exactly
the Whitham evolution of the functional $k ([{\bf U}], W)$
corresponding to the $\nu$-flow of the KdV hierarchy. So we
have

$$\{ k (W) , I^{\nu} \}|_{\cal K} \,\, = \,\,
\epsilon \left( \omega^{\nu} (k, A, n) \, + \, \sum_{l\geq1}
\epsilon^{2l} \, \omega^{\nu}_{(2l)} (k, A, n, \dots)
\right)_{W} \,\, =$$
$$= \,\, \epsilon \left( \omega^{\nu}_{0} ({\bf U}) \, + \, 
\sum_{l\geq1} \epsilon^{2l} \, {\tilde{\omega}}^{\nu}_{(2l)}
({\bf U}, {\bf U}_{X}, \dots) \right)_{W} $$
with some functionals 
${\tilde{\omega}}^{\nu}_{(2l)} ({\bf U}, {\bf U}_{X}, \dots)$
according to (\ref{kAnTnu}).

 According to the structure of the bracket 
$\{ k (W) , U^{\nu} (Y) \}|_{\cal K}$ we should have then

$$\{ k (W) , U^{\nu} (Y) \}|_{\cal K} \,\, = \,\,
\epsilon \left( \omega^{\nu}_{0} ({\bf U}(W)) \, + \,
\sum_{l\geq1} \epsilon^{2l} \, {\tilde{\omega}}^{\nu}_{(2l)}
({\bf U}, {\bf U}_{W}, \dots) \right)_{W} \, \delta (W -Y) 
\, + $$
$$+ \, \sum_{s\geq1} \epsilon^{s} \, \kappa^{\nu L}_{(s)}
({\bf U}, {\bf U}_{W}, \dots) \, \delta^{(s)} (W -Y) $$
where $\kappa^{\nu L}_{(s)}$ are some local functionals of
$({\bf U}, {\bf U}_{W}, \dots)$ given by sums of terms of degree
$\geq 0$.

 We can write then

$${1 \over 2 \epsilon} \, \int_{-\infty}^{+\infty} d W \,
\psi_{\theta} (\theta , [{\bf U}], X ) \,\, {\rm sgn} (X - W)
\,\, \{ k (W) , U^{\nu} (Y) \}|_{\cal K} \,\, = $$
$$= \, \psi_{\theta} (\theta, [{\bf U}], X)
\left( \omega^{\nu}_{0} ({\bf U}(X)) \, + \,
\sum_{l\geq1} \epsilon^{2l} \, {\tilde{\omega}}^{\nu}_{(2l)}
({\bf U}, {\bf U}_{X}, \dots) \right) \, \delta (X -Y) \, -$$
$$-  {1 \over 2} \int_{-\infty}^{+\infty} \!\! d W \,
\psi_{\theta} (\theta, [{\bf U}], X) \, {\rm sgn} (X - W) 
\left( \omega^{\nu}_{0} ({\bf U}(W)) + 
\sum_{l\geq1} \epsilon^{2l} \, {\tilde{\omega}}^{\nu}_{(2l)}
({\bf U}, {\bf U}_{W}, \dots) \right)  \delta^{\prime} (W -Y) + $$
$$+ \, {1 \over 2 \epsilon} \, \int_{-\infty}^{+\infty} d W \,
\psi_{\theta} (\theta , [{\bf U}], X ) \,\, {\rm sgn} (X - W)
\,\, \left( \sum_{s\geq1} \epsilon^{s} \, \kappa^{\nu L}_{(s)}
({\bf U}, {\bf U}_{W}, \dots) \, \delta^{(s)} (W -Y) \right) $$

 In the same way we obtain

$$\{ \varphi (\theta, X) , I^{\nu} \}|_{\cal K} \,\, = \,\,
\psi_{\theta} \left( {S (X) \over \epsilon} + \theta, 
[{\bf U}], X) \right) \left(\omega^{\nu}_{0} ({\bf U}(X)) +
\sum_{l\geq1} \epsilon^{2l} \, {\tilde{\omega}}^{\nu}_{(2l)} 
({\bf U}, {\bf U}_{X}, \dots) \right) \, + $$
$$+ \, \int_{-\infty}^{+\infty} d W \,
\psi_{U^{\mu}} \left( {S (X) \over \epsilon} + \theta,
[{\bf U}], X, W \right) 
\{ U^{\mu} (W) , I^{\nu} \}|_{\cal K} $$
where

$$\psi_{U^{\mu}} (\theta, [{\bf U}], X, W ) \,\, \equiv \,\,
{\delta \psi (\theta , [{\bf U}], X ) \over \delta U^{\mu} (W)} $$

 So, from the structure of the bracket 
$\{ \varphi (\theta, X) , U^{\nu} (Y) \}|_{\cal K}$
we can conclude

$$\chi^{\nu}_{L} \left( {S (X) \over \epsilon} + \theta, X, Y
\right) \,\, \equiv \,\, 
\{ \varphi (\theta, X) , U^{\nu} (Y) \}|_{\cal K} \,\, = $$
$$= \, \psi_{\theta} \left( {S (X) \over \epsilon} + \theta,
[{\bf U}], X \right) \left(\omega^{\nu}_{0} ({\bf U}(X)) +
\sum_{l\geq1} \epsilon^{2l} \, {\tilde{\omega}}^{\nu}_{(2l)}
({\bf U}, {\bf U}_{X}, \dots) \right) \delta (X -Y) \, + $$
\begin{equation}
\label{phiUbrack}
+ \, \int_{-\infty}^{+\infty} d W \,
\psi_{U^{\mu}} \left( {S (X) \over \epsilon} + \theta,
[{\bf U}], X, W \right)
\{ U^{\mu} (W) , U^{\nu} (Y) \}|_{\cal K} \, + 
\end{equation}
$$+ \, \sum_{s\geq1} \epsilon^{s} \, \lambda^{\nu L}_{(s)}
\left( {S (X) \over \epsilon} + \theta, [{\bf U}], X \right)
\delta^{(s)} (X - Y) $$
for some local functionals 
$\lambda^{\nu L}_{(s)} (\theta, [{\bf U}], X)$ on ${\cal K}$,
polynomial in the derivatives 
$({\bf U}_{X}, {\bf U}_{XX}, \dots )$ and given by sums of
terms of degree $\geq 0$.

 In the same way we put

$$\chi^{\nu}_{R} \left( Y, X, {S (X) \over \epsilon} + \theta, 
\right) \,\, \equiv \,\, 
\{ U^{\nu} (Y) , \varphi (\theta, X) \}|_{\cal K} \, = $$
$$= \, - \, \delta (Y - X)
\left(\omega^{\nu}_{0} ({\bf U}(X)) +
\sum_{l\geq1} \epsilon^{2l} \, {\tilde{\omega}}^{\nu}_{(2l)}
({\bf U}, {\bf U}_{X}, \dots) \right)
\psi_{\theta} \left( {S (X) \over \epsilon} + \theta,
[{\bf U}], X \right) + $$
\begin{equation}
\label{chiRnu}
+ \, \int_{-\infty}^{+\infty} d W \,
\{ U^{\nu} (Y) , U^{\mu} (W) \}|_{\cal K} \, 
\psi_{U^{\mu}} \left( {S (X) \over \epsilon} + \theta,
[{\bf U}], X, W \right) + 
\end{equation}
$$+ \, \sum_{s\geq1} \epsilon^{s} \, \delta^{(s)} (Y - X)
\, \lambda^{\nu R}_{(s)} \left( {S (X) \over \epsilon} + \theta, 
[{\bf U}], X \right) $$

 Let us denote also

$$\zeta_{L} \left( {S (X) \over \epsilon} + \theta,
X, Y, \right) \,\, \equiv \,\, 
\{ \varphi (\theta, X) , k (Y) \}|_{\cal K} $$
$$\zeta_{R} \left( Y, X, {S (X) \over \epsilon} + \theta,
\right) \,\, \equiv \,\,
\{ k (Y) , \varphi (\theta, X) \}|_{\cal K} $$

 We have now

$$\alpha^{\nu} \left( {S (X) \over \epsilon} + \theta, X, Y
\right) \,\, \equiv \,\, 
\{g (\theta, X) , U^{\nu} (Y) \}|_{\cal K} \,\, =$$
$$= \, \{\varphi (\theta, X) , U^{\nu} (Y) \}|_{\cal K} \, - \,
\int_{-\infty}^{+\infty} d W \,
\psi_{U^{\mu}} \left( {S (X) \over \epsilon} + \theta,
[{\bf U}], X, W \right)
\{ U^{\mu} (W) , U^{\nu} (Y) \}|_{\cal K} \, - $$
$$- \, {1 \over 2 \epsilon} \, \int_{-\infty}^{+\infty} d W \,
\psi_{\theta} \left({S (X) \over \epsilon} + \theta , 
[{\bf U}], X \right) \,\, {\rm sgn} (X - W) \,\, 
\{ k (W) , U^{\nu} (Y) \}|_{\cal K} \, =$$

$$= \, \sum_{s\geq1} \epsilon^{s} \, \lambda^{\nu L}_{(s)}
\left( {S (X) \over \epsilon} + \theta, [{\bf U}], X \right)
\delta^{(s)} (X - Y) \, +$$
$$+ {1 \over 2} \int_{-\infty}^{+\infty} \!\!\! d W \,
\psi_{\theta} \left({S (X) \over \epsilon} + \theta , 
[{\bf U}], X \right) \, {\rm sgn} (X - W)
\left( \omega^{\nu}_{0} (W) +
\sum_{l\geq1} \epsilon^{2l} {\tilde{\omega}}^{\nu}_{(2l)}
(W) \right)  \delta^{\prime} (W -Y) - $$
$$- \, {1 \over 2 \epsilon} \, \int_{-\infty}^{+\infty} d W \,
\psi_{\theta} \left({S (X) \over \epsilon} + \theta , 
[{\bf U}], X \right) \,\, {\rm sgn} (X - W)
\,\, \left( \sum_{s\geq1} \epsilon^{s} \, \kappa^{\nu L}_{(s)}
(W) \, \delta^{(s)} (W -Y) \right) $$

 Using the same arguments we obtain that for the case of the 
Gardner - Zakharov - Faddeev bracket we have the following 
equation for the functions $\beta^{\nu} (\theta, Z, Y)$:

\begin{equation}
\label{uravnbeta}
\int_{-\infty}^{+\infty} \int_{0}^{2\pi} 
L (\theta, \theta^{\prime}, X, Z) \,\, 
\beta^{\nu} (\theta^{\prime}, Z, Y) \,\, 
{d \theta^{\prime} \over 2\pi}
\, dZ \,\, = \,\, \alpha^{\nu} (\theta, X, Y) 
\end{equation}
where

$$L (\theta, \theta^{\prime}, X, Z) \,\, = \,\,
k \, \delta^{\prime} (\theta - \theta^{\prime}) \,
\delta (X - Z) \,\, + \,\, 
\epsilon \, \delta (\theta - \theta^{\prime}) \,
\delta^{\prime} (X - Z) \, - $$
$$- \, \int_{-\infty}^{+\infty} d W \, 
\chi^{\mu}_{L} (\theta, X, W) \,
{\delta \psi (\theta^{\prime}, [{\bf U}], Z ) \over 
\delta U^{\mu} (W)} \, - \, \int_{-\infty}^{+\infty} d W \,
{\delta \psi (\theta , [{\bf U}], X ) \over \delta U^{\mu} (W)} \,
\chi^{\mu}_{R} (W, Z, \theta^{\prime}) \, + $$
$$+ \, \int_{-\infty}^{+\infty} \int_{-\infty}^{+\infty} d W\,dV\,
{\delta \psi (\theta , [{\bf U}], X ) \over \delta U^{\mu} (W)} \, 
\{ U^{\mu} (W) , U^{\gamma} (V) \}|_{\cal K}
{\delta \psi (\theta^{\prime}, [{\bf U}], Z ) \over 
\delta U^{\gamma} (V)} \, - $$
$$- \, {1 \over 2 \epsilon} \, \int_{-\infty}^{+\infty} d W \,
\zeta_{L} (\theta, X, W) \,\, {\rm sgn} \, (Z - W) \,\, 
\psi_{\theta^{\prime}} (\theta^{\prime}, [{\bf U}], Z ) \, - $$
$$- \, {1 \over 2 \epsilon} \, \int_{-\infty}^{+\infty} d W \,
\psi_{\theta} (\theta , [{\bf U}], X ) \,\, {\rm sgn} \, (X - W) 
\,\, \zeta_{R} (W, Z, \theta^{\prime}) \,\, + $$
$$+ \, {1 \over 2 \epsilon} \, \int_{-\infty}^{+\infty}
\int_{-\infty}^{+\infty} d W \, d V \,  
{\delta \psi (\theta , [{\bf U}], X ) \over \delta U^{\mu} (W)} \,
\{ U^{\mu} (W) , k (V) \}|_{\cal K} \,\, {\rm sgn} \, (Z - V) 
\,\, \psi_{\theta^{\prime}} (\theta^{\prime}, [{\bf U}], Z ) 
\, + $$
$$+ \, {1 \over 2 \epsilon} \, \int_{-\infty}^{+\infty}
\int_{-\infty}^{+\infty} d W \, d V \,
\psi_{\theta} (\theta , [{\bf U}], X ) \,\, {\rm sgn} \, (X - W)
\,\, \{ k (W) , U^{\gamma} (V) \}|_{\cal K} 
{\delta \psi (\theta^{\prime}, [{\bf U}], Z ) \over
\delta U^{\gamma} (V)} \, + $$
$$+ {1 \over 4 \epsilon^{2}} \int_{-\infty}^{+\infty} \!\!
\int_{-\infty}^{+\infty} \!\!\! d W d V \,
\psi_{\theta} (\theta , [{\bf U}], X ) \, {\rm sgn} (X - W)
\, \{ k (W) , k (V) \}|_{\cal K} \,
{\rm sgn} (Z - V) \,
\psi_{\theta^{\prime}} (\theta^{\prime}, [{\bf U}], Z ) $$

 Let us note now that the bracket 
$\{ U^{\nu} (X) , U^{\mu} (Y) \}|_{\cal K}$ has the order
${\cal O}(\epsilon)$ and, besides that, it's main term in the
$\epsilon$-expansion coincides precisely with the
Dubrovin - Novikov bracket defined above.

 Let us put now the additional condition

\begin{equation}
\label{psithetaort}
\int_{0}^{2\pi} \psi_{\theta} (\theta , [{\bf U}], Z ) \,\,
\beta^{\nu} (\theta, Z, Y) \,\, {d \theta \over 2\pi} 
\,\, \equiv \,\, 0 
\end{equation}
which will be confirmed {\it aposteriori} for our
$\beta^{\nu} (\theta, Z, Y)$. We can reduce then the operator 
$L (\theta, \theta^{\prime}, X, Z)$ to the form

$$L_{eff} (\theta, \theta^{\prime}, X, Z) \,\, = \,\,
k \, \delta^{\prime} (\theta - \theta^{\prime}) \, 
\delta (X - Z) \,\, + \,\,
\epsilon \, \delta (\theta - \theta^{\prime}) \,
\delta^{\prime} (X - Z) \,\, - $$
$$- \, \int_{-\infty}^{+\infty} d W \,
\chi^{\mu}_{L} (\theta, X, W) \,
{\delta \psi (\theta^{\prime}, [{\bf U}], Z ) \over
\delta U^{\mu} (W)} \, - \, \int_{-\infty}^{+\infty} d W \,
{\delta \psi (\theta , [{\bf U}], X ) \over \delta U^{\mu} (W)} \,
\chi^{\mu}_{R} (W, Z, \theta^{\prime}) \, + $$
$$+ \, \int_{-\infty}^{+\infty} \int_{-\infty}^{+\infty} d W\,dV\,
{\delta \psi (\theta , [{\bf U}], X ) \over \delta U^{\mu} (W)} \,
\{ U^{\mu} (W) , U^{\gamma} (V) \}|_{\cal K}
{\delta \psi (\theta^{\prime}, [{\bf U}], Z ) \over
\delta U^{\gamma} (V)} \, - $$
$$- \, {1 \over 2 \epsilon} \, \int_{-\infty}^{+\infty} d W \,
\psi_{\theta} (\theta , [{\bf U}], X ) \,\, {\rm sgn} \, (X - W)
\,\, \zeta_{R} (W, Z, \theta^{\prime}) \,\, + $$
$$+ \, {1 \over 2 \epsilon} \, \int_{-\infty}^{+\infty}
\int_{-\infty}^{+\infty} d W \, d V \,
\psi_{\theta} (\theta , [{\bf U}], X ) \,\, {\rm sgn} \, (X - W)
\,\, \{ k (W) , U^{\gamma} (V) \}|_{\cal K}
{\delta \psi (\theta^{\prime}, [{\bf U}], Z ) \over
\delta U^{\gamma} (V)} $$

 Let us define now the functions 
$\beta^{\nu} (\theta, Z, Y)$ as the solutions of the equations

\begin{equation}
\label{Isyst}
\int_{-\infty}^{+\infty} \int_{0}^{2\pi} 
L^{I}_{eff} (\theta, \theta^{\prime}, X, Z) \,\,
\beta^{\nu} (\theta^{\prime}, Z, Y) \,\,
{d \theta^{\prime} \over 2\pi}
\, dZ \,\, = \,\, \alpha^{\nu I} (\theta, X, Y)
\end{equation}
where

$$L_{eff} (\theta, \theta^{\prime}, X, Z) \,\, = \,\,
L^{I}_{eff} (\theta, \theta^{\prime}, X, Z)  \, + \,
L^{II}_{eff} (\theta, \theta^{\prime}, X, Z) $$
$$\alpha^{\nu} (\theta, X, Y) \,\, = \,\,
\alpha^{\nu I} (\theta, X, Y) \, + \,
\alpha^{\nu II} (\theta, X, Y) $$
and

$$L^{I}_{eff} (\theta, \theta^{\prime}, X, Z) \,\, = \,\,
k \, \delta^{\prime} (\theta - \theta^{\prime}) \,
\delta (X - Z) \,\, + \,\,
\epsilon \, \delta (\theta - \theta^{\prime}) \,
\delta^{\prime} (X - Z) \,\, - $$
$$- \, \int_{-\infty}^{+\infty} d W \,
\chi^{\mu}_{L} (\theta, X, W) \,
{\delta \psi (\theta^{\prime}, [{\bf U}], Z ) \over
\delta U^{\mu} (W)} \, - \, \int_{-\infty}^{+\infty} d W \,
{\delta \psi (\theta , [{\bf U}], X ) \over \delta U^{\mu} (W)} \,
\chi^{\mu}_{R} (W, Z, \theta^{\prime}) \, + $$ 
$$+ \, \psi_{\theta} (\theta , [{\bf U}], X ) 
\left( \omega^{\mu}_{0} (X) +
\sum_{l\geq1} \epsilon^{2l} {\tilde{\omega}}^{\mu}_{(2l)}
(X) \right) {\delta \psi (\theta^{\prime}, [{\bf U}], Z ) 
\over \delta U^{\mu} (X)} \, +$$ 
$$+ \, \int_{-\infty}^{+\infty} \int_{-\infty}^{+\infty} d W\,dV\,
{\delta \psi (\theta , [{\bf U}], X ) \over \delta U^{\mu} (W)} \,
\{ U^{\mu} (W) , U^{\gamma} (V) \}|_{\cal K}
{\delta \psi (\theta^{\prime}, [{\bf U}], Z ) \over  
\delta U^{\gamma} (V)} \,\, = $$
$$= \,\, k \, \delta^{\prime} (\theta - \theta^{\prime}) \,
\delta (X - Z) \,\, + \,\,
\epsilon \, \delta (\theta - \theta^{\prime}) \,
\delta^{\prime} (X - Z) \,\, - $$
$$- \,\, \sum_{s\geq1} \epsilon^{s} \,
\lambda^{\mu L}_{(s)} (\theta, [{\bf U}], X) \,
{d^{s} \over d X^{s}} 
{\delta \psi (\theta^{\prime}, [{\bf U}], Z) \over
\delta U^{\mu} (X)} \, - \,
\int_{-\infty}^{+\infty} d W \,
{\delta \psi (\theta , [{\bf U}], X ) \over \delta U^{\mu} (W)} \,
\chi^{\mu}_{R} (W, Z, \theta^{\prime}) $$

$$L^{II}_{eff} (\theta, \theta^{\prime}, X, Z) \, = \,
- \, {1 \over 2 \epsilon} \, \int_{-\infty}^{+\infty} d W \,
\psi_{\theta} (\theta , [{\bf U}], X ) \, {\rm sgn} \, (X - W)
\, \zeta_{R} (W, Z, \theta^{\prime}) \, - $$
$$- \, {1 \over 2} \int_{-\infty}^{+\infty} d W \, 
\psi_{\theta} (\theta , [{\bf U}], X ) \, {\rm sgn} \, (X - W)
\, \left( \omega^{\mu}_{0} (W) + \sum_{l\geq1} 
\epsilon^{2l} {\tilde{\omega}}^{\mu}_{(2l)} (W) \right)
{d \over d W} {\delta \psi (\theta^{\prime}, [{\bf U}], Z ) 
\over \delta U^{\mu} (W)} \, + $$
$$+ {1 \over 2 \epsilon} \int_{-\infty}^{+\infty} d W \,
\psi_{\theta} (\theta , [{\bf U}], X ) \, {\rm sgn} \, (X - W)
\, \sum_{s\geq1} \epsilon^{s} \kappa^{\mu L}_{(s)} (W)
{d^{s} \over d W^{s}} {\delta \psi (\theta^{\prime}, [{\bf U}], Z )
\over \delta U^{\mu} (W)} $$

$$\alpha^{\nu I} (\theta, X, Y) \,\, = \,\,
\sum_{s\geq1} \epsilon^{s} \, \lambda^{\nu L}_{(s)}
( \theta, [{\bf U}], X ) \, \delta^{(s)} (X - Y) $$

$$\alpha^{\nu II} (\theta, X, Y)  \,\, = $$
$$= \,{1 \over 2} \int_{-\infty}^{+\infty} \!\!\! d W \,
\psi_{\theta} (\theta , [{\bf U}], X) \, {\rm sgn} (X - W)
\left( \omega^{\nu}_{0} (W) +
\sum_{l\geq1} \epsilon^{2l} {\tilde{\omega}}^{\nu}_{(2l)}
(W) \right)  \delta^{\prime} (W -Y) - $$
$$- \, {1 \over 2 \epsilon} \, \int_{-\infty}^{+\infty} d W \,
\psi_{\theta} (\theta , [{\bf U}], X ) \, {\rm sgn} (X - W)
\, \sum_{s\geq1} \epsilon^{s} \, \kappa^{\nu L}_{(s)}
(W) \, \delta^{(s)} (W -Y) $$

 We have to prove now that the solutions 
$\beta^{\nu} (\theta, Z, Y)$ satisfy in fact system
(\ref{uravnbeta}) and have the form represented by 
(\ref{beta1def})-(\ref{beta2def}). So, let us discuss first
the resolvability of system (\ref{Isyst}). According to relations
(\ref{phiUbrack})-(\ref{chiRnu})
we can write the main part (in $\epsilon$)
of the operator ${\hat L}^{I}_{eff}$ in the form:

$$L^{I}_{eff (0)} (\theta, \theta^{\prime}, X, Z) \, = \,
k \, \delta^{\prime} (\theta - \theta^{\prime}) \,
\delta (X - Z) \, + \, \Phi_{U^{\nu}} (\theta, {\bf U}(X)) \,
\omega^{\nu}_{0}(X) \,
\Phi_{\theta^{\prime}} (\theta^{\prime}, {\bf U}(X)) \,
\delta (X - Z) $$

 The operator ${\hat L}^{I}_{eff (0)}$ gives a set of
independent operators at different $X$ where the operator

$$k \, \delta^{\prime} (\theta - \theta^{\prime}) \, + \,
\Phi_{U^{\nu}} (\theta, {\bf U}(X)) \, \omega^{\nu}_{0}(X) \,
\Phi_{\theta^{\prime}} (\theta^{\prime}, {\bf U} (X)) $$
has at every $X$
exactly two linearly independent left eigen-vectors on the
space of periodic functions in $\theta$

$${\eta}_{1} (\theta, X) \,\, = \,\, 1 \,\,\,\,\, , \,\,\,\,\,
{\eta}_{2} (\theta, X) \,\, = \,\, \Phi (\theta, {\bf U}(X)) $$
corresponding to the zero eigen-values.

 The vectors ${\eta}_{1} (\theta, X) \, \delta (V - X)$ and
${\eta}_{2} (\theta, X) \, \delta (V - X)$ give the main parts
of the vectors

\begin{equation}
\label{vectorsI}
{\delta U^{0}(V) \over \delta \varphi (\theta, X)}|_{\cal K}
\,\, = \,\, \delta (V - X) \,\,\,\,\, , \,\,\,\,\, 
{\delta U^{1}(V) \over \delta \varphi (\theta, X)}|_{\cal K}
\,\, = \,\, \psi (\theta , [{\bf U}], X ) \, \delta (V - X) 
\end{equation}
which are the left eigen-vectors of the operator 
${\hat L}^{I}_{eff}$ corresponding to the zero eigen-values.

 It's not difficult to see now that the orthogonality of the
values $\{ g (\theta, X) , U^{\nu}(Y) \}$ to vectors
(\ref{vectorsI}) on ${\cal K}$ implies the orthogonality of 
$\alpha^{\nu I} (\theta, X, Y)$
to the same vectors. So, we get that system 
(\ref{Isyst}) is a compatible system which can be resolved
recursively in all the orders of $\epsilon$. The right-hand 
part of system (\ref{Isyst}) has the form analogous to
(\ref{beta1def})-(\ref{beta2def}) so it's not difficult to
see that all the $\beta^{\nu} (\theta, Z, Y)$ have the
necessary form in this case. Using also the fact 
$\alpha^{\nu I} (\theta, X, Y) = {\cal O}(\epsilon)$
we get that the solutions $\beta^{\nu} (\theta, Z, Y)$
have exactly the required form (\ref{beta1def})-(\ref{beta2def})
being written as formal series in $\epsilon$. Besides that,
condition (\ref{psithetaort}) can be also derived from a not
complicated analysis of system (\ref{Isyst}) by use of the
same left eigen-vectors of ${\hat L}^{I}_{eff}$ corresponding
to the zero eigen-values.

 Finally, let us prove the relation

\begin{equation}
\label{Leffrel}
\int_{-\infty}^{+\infty} \int_{0}^{2\pi}
L_{eff} (\theta, \theta^{\prime}, X, Z) \,\,
\beta^{\nu} (\theta^{\prime}, Z, Y) \,\,
{d \theta^{\prime} \over 2\pi}
\, dZ \,\, = \,\, \alpha^{\nu} (\theta, X, Y)
\end{equation}
for the $\beta^{\nu} (\theta, Z, Y)$ found from (\ref{Isyst}).

 Let us note that the difference in the images of the operators
${\hat L}_{eff}$ and ${\hat L}^{I}_{eff}$ for our
$\beta^{\nu} (\theta, Z, Y)$ is proportional to
$\psi_{\theta} (\theta , [{\bf U}], X )$ at every $(X, Y)$.
The same is also valid for $\alpha^{\nu II} (\theta, X, Y)$
which is the difference between 
$\alpha^{\nu} (\theta,X,Y)$ and $\alpha^{\nu I} (\theta,X,Y)$.
It's not difficult to check then that relation (\ref{Leffrel})
follows from (\ref{Isyst}) and the orthogonality of the values

$$\int_{-\infty}^{+\infty} \int_{0}^{2\pi}
L^{II}_{eff} (\theta, \theta^{\prime}, X, Z) \,\,
\beta^{\nu} (\theta^{\prime}, Z, Y) \,\,
{d \theta^{\prime} \over 2\pi}
\, dZ \,\, - \,\, \alpha^{\nu II} (\theta, X, Y) $$
to the vectors $\delta U^{2} (V) / \delta \varphi (\theta, X)$
which takes place for our $\beta^{\nu} (\theta, Z, Y)$.

 Using formula (\ref{DiracBracket}) we can claim now that the
restricted Poisson bracket \linebreak
$\{ U^{\nu}(X) , U^{\mu}(Y) \}_{D}$
has exactly form (\ref{DeformBracket}).

 Let us recall now the the functionals 
$I^{\nu} = \int_{-\infty}^{+\infty} U^{\nu} (X) d X$
leave invariant the submanifold ${\cal K}$ as was proved in
Theorem 4.1. This means in particular that the flows generated
by $I^{\nu}$ on ${\cal K}$ coincide with their flows generated
in the Dirac Poisson structure on this submanifold. Thus, we 
obtain that the functionals $I^{\nu}$ play the role of the
Hamiltonian functions for the higher deformed Whitham systems
(\ref{kAnTnu}) and, in particular, the functional $I^{2}$
plays the role of the Hamiltonian function for the deformed
Whitham system (\ref{epskan}) after the restriction on
${\cal K}$. In the same way the functionals $I^{0}$ and $I^{1}$
play the role of the annihilator and the momentum functional
for the restricted Gardner - Zakharov - Faddeev bracket
respectively.

 At last, let us say that the proof of the Theorem for the case
of the Magri bracket repeats completely the proof for the
Gardner - Zakharov - Faddeev case.

{\hfill Theorem is proved.}

\vspace{0.5cm}

{\bf Remark 4.1.}

 It's not difficult to see that the main $(\sim \epsilon)$
term of the Dirac bracket $\{ U^{\nu}(X) , U^{\mu}(Y) \}_{D}$
on ${\cal K}$ coincides with the Dubrovin - Novikov bracket
for the Whitham system given by the "averaging procedure"
described above. The Dubrovin - Novikov bracket obtained
from the Gardner - Zakharov - Faddeev bracket and the Magri 
bracket respectively are compatible with each other and give
a bi-Hamiltonian structure for the pure Whitham system for KdV.
However, we can not claim here the same property for the case
of the Dirac brackets obtained as the restrictions of the
Gardner - Zakharov - Faddeev bracket and the Magri bracket
on ${\cal K}$ since the Dirac procedure does not preserve
the compatibility of the brackets in general case. 

\vspace{0.5cm}

{\bf Remark 4.2.}

 It's not difficult to see that the functional

$$\int_{-\infty}^{+\infty} k (X) \,\, d X $$
plays the role of annihilator for the restricted Poisson 
brackets both in the cases of the Gardner - Zakharov - Faddeev
bracket and the Magri bracket. This circumstance is connected
with the conservation of the value $S (+\infty) - S (-\infty)$
by the flows generated by the "modified" functionals

$$\int_{-\infty}^{+\infty} {\tilde{U}}^{\nu} (X) 
\,\, q (X) \,\, d X $$
with $q (X)$ having compact support and has a general
character for the restricted field-theoretical Poisson brackets.

\section{Some remarks on the averaging of the Lagrangian
structures.}
\setcounter{equation}{0}

 At the end let us discuss also the averaging
of Lagrangian functions for the deformed Whitham systems. We
will restrict ourselves here only to the situation of the
local Lagrangian functions which was considered first by
Whitham (\cite{whith1,whith2,whith3}) in connection with the
pure Whitham approach.

 As it is well-known the Gardner - Zakharov - Faddeev bracket
corresponds to the local Lagrangian formalism of the KdV
equation (\ref{ekdv}) having the form

\begin{equation}
\label{KdVLagr}
{\delta \over \delta v(X,T)} \int\int
\left[- v_{X} v_{T} \, - \,
{\epsilon \over 3} \, v_{X}^{3} \, + \, 
\epsilon^{2} \, v_{XX}^{2} \right]
dX \, dT \,\,\, = \,\,\, 0
\end{equation}  
(where $\varphi = \epsilon \, v_{X}$) which gives

\begin{equation}
\label{vKdV}  
v_{XT} \, + \, \epsilon v_{X} v_{XX} \, + \, 
\epsilon^{2} \, v_{XXXX} \,\,\, = \,\,\, 0   
\end{equation}

 We introduce also the Whitham pseudo-phase $\Sigma(X,T)$ and
look for the solution of (\ref{vKdV}) having the form

\begin{equation}
\label{vsolution}
v(\theta, X,T) \,\,\, = \,\,\, V^{(tot)} \left(
{S(X,T) \over \epsilon} + \theta, X, T \right)
\,\, + \,\, {\Sigma(X,T) \over \epsilon} 
\end{equation}

 We require now that $V^{(tot)}(\theta, X, T)$ is a periodic
function in $\theta$ having the form

\begin{equation}
\label{vsol}
V^{(tot)}(\theta, X, T) \,\,\, = \,\,\,
\sum_{k\geq0} V_{(k)} (\theta, X, T)
\end{equation}
where all the functions $V_{(k)} (\theta, X, T)$ are 
local functionals of
$$(k = S_{X}, S_{T}, n = \Sigma_{X}, k_{X},
S_{TX}, n_{X}, k_{XX}, S_{TXX}, n_{XX}, \dots)$$
having degree $k$ according to the gradation rule:

\vspace{0.5cm}

1) All the functions $f(k, S_{T}, n)$ have degree $0$;
 
2) The derivatives $k_{kX}$, $S_{TkX}$, $n_{kX}$
have degree $k$;

3) The degree of the product of functions having certain
degrees is equal to the sum of their degrees.

\vspace{0.5cm}

 According to the normalization of $\Sigma(X,T)$ we put the 
conditions

\begin{equation}
\label{V0norm}
\int_{0}^{2\pi} V_{(k)} (\theta, X, T) \,\, 
{d \theta \over 2\pi} \,\, \equiv \,\, 0
\end{equation}
for all $V_{(k)} (\theta, X, T)$.

 Let us note that the choice of the parameters 
$(k, S_{T}, n)$ instead of $(k, A, n)$ is more convenient
for the consideration of the Lagrangian structures in our
approach. We remind also that the expression for $S_{T}$
is given by relation (\ref{epsdisp}).

 Easy to see that the form (\ref{vsolution}) gives the form of
$\phi (\theta, X, T)$ we consider and all the functions
$V_{(k)} (\theta, X, T)$ are uniquely defined by the terms of
series (\ref{phiepsexp}). Indeed, let us first re-expand series
(\ref{phiepsexp}) according to the new gradation rule, i.e.

$$\phi (\theta, X, T) \,\, = \,\, \sum_{l\geq0} \epsilon^{l} \,
\Phi^{\prime}_{(l)} \left( {S (X,T) \over \epsilon} + \theta,
X, T \right) $$
where all $\Phi^{\prime}_{(l)}$ have degree $l$ according to 
the rules formulated above.

 Then we have

$$k \, V_{(l)\theta} (\theta, X, T) \, + \,
V_{(l-1)X} (\theta, X, T) \,\, = \,\, \Phi^{\prime}_{(l)}
(\theta, X, T) \,\,\,\,\, , \,\,\,\,\, l \geq 1$$
which defines uniquely all $V_{(l)} (\theta, X, T)$ view
normalization rule (\ref{V0norm}) and we have

$$\Sigma_{X} (X, T) \,\, = \,\, \int_{0}^{2\pi}
\phi (\theta, X, T) \,\, {d \theta \over 2\pi} $$

 Finally, we can substitute series (\ref{vsolution}) in the
Lagrangian principle

$$\delta \int\int\int_{0}^{2\pi} {\cal L}(\theta, X, T) \,\,
{d \theta \over 2\pi} \, dX \, dT $$
with the Lagrangian density

$${\cal L} \,\, = \,\,
- S_{X} S_{T} \left( V^{(tot)}_{\theta} \right)^{2}
\, - \, \Sigma_{X} \Sigma_{T} \, - \, {1 \over 3}
S_{X}^{3} \left( V^{(tot)}_{\theta} \right)^{3} \, - \,
\Sigma_{X} S_{X}^{2} \left( V^{(tot)}_{\theta} \right)^{2}
\, - \, {1 \over 3} \Sigma_{X}^{3} \, + \, S_{X}^{4}
\left( V^{(tot)}_{\theta\theta} \right)^{2} \,\, + $$

\vspace{0.3cm}

$$+ \,\, \epsilon \left(
-  \, S_{X}  V^{(tot)}_{\theta} V^{(tot)}_{T}
\, - \, S_{T}  V^{(tot)}_{\theta} V^{(tot)}_{X}
\, - \, \Sigma_{T} V^{(tot)}_{X} \, - \,
\Sigma_{X} V^{(tot)}_{T} \,\, - \right. $$
$$\left. - \, S_{X}^{2} 
\left( V^{(tot)}_{\theta} \right)^{2}
V^{(tot)}_{X} \, - \, 2 \Sigma_{X} S_{X} V^{(tot)}_{\theta}
V^{(tot)}_{X} \, - \, \Sigma_{X}^{2} V^{(tot)}_{X} \, + \,
S_{X}^{3} V^{(tot)}_{\theta\theta} V^{(tot)}_{\theta X}
\right) \, + $$

\vspace{0.3cm}
$$+ \,\, \epsilon^{2} \left(
- \, V^{(tot)}_{T} V^{(tot)}_{X} \, - \,
S_{X} V^{(tot)}_{\theta} \left( V^{(tot)}_{X} \right)^{2}
\, - \, \Sigma_{X} \left( V^{(tot)}_{X} \right)^{2} \, + \,
4 S_{X}^{2} \left( V^{(tot)}_{\theta X} \right)^{2} \, + 
\right. $$
$$\left. + \, S_{XX}^{2} \left( V^{(tot)}_{\theta} \right)^{2} 
\, + \, S_{X}^{2} V^{(tot)}_{\theta\theta} V^{(tot)}_{XX} \, + \,
4 S_{X} S_{XX} V^{(tot)}_{\theta} V^{(tot)}_{\theta X} \, + \,
\Sigma_{XX}^{2} \right) \, +$$

\vspace{0.3cm}

$$+ \,\, \epsilon^{3} \left( - \, {1 \over 3}
\, \left( V^{(tot)}_{X} \right)^{3} \, + \,
4 S_{X} V^{(tot)}_{\theta X} V^{(tot)}_{XX} \, + \,
2 S_{XX} V^{(tot)}_{\theta} V^{(tot)}_{XX} \, + \,
2 \Sigma_{XX} V^{(tot)}_{XX} \right) \, + \, \epsilon^{4} 
\left( V^{(tot)}_{XX} \right)^{2}$$

\vspace{0.3cm}

 The averaged Lagrangian function

$$\langle {\cal L} \rangle (X, T) \,\, \equiv \,\,
\int_{0}^{2\pi} {\cal L} (\theta, X, T) \,\, 
{d \theta \over 2\pi} $$
can be also represented in the graded form with respect to
the parameters $(k = S_{X}, S_{T}, n = \Sigma_{X})$ and the
Lagrangian equations

\begin{equation}
\label{LagrAv}
{\delta \over \delta S (X, T)} \int_{-\infty}^{+\infty}
\int_{-\infty}^{+\infty} \langle {\cal L} \rangle (X, T) \,
d X dT \,\,\,\,\, , \,\,\,\,\,
{\delta \over \delta \Sigma (X, T)} \int_{-\infty}^{+\infty}
\int_{-\infty}^{+\infty} \langle {\cal L} \rangle (X, T) \,
d X dT
\end{equation}
give a system equivalent to (\ref{epskan}).

 The Hamiltonian formalism in the parameters 
$(k, S_{T}, n)$ can be written using Lagrangian formalism
(\ref{LagrAv}). We get then the Poisson bracket in the
canonical form:

\begin{equation}
\label{CanHamStr}
\{ n (X) , n (Y) \} \,\, = \,\, \delta^{\prime} (X - Y) 
\,\,\,\,\,\,\,\, , \,\,\,\,\,\,\,\,
\{ k (X) , J (Y) \} \,\, = \,\, \delta^{\prime} (X - Y)
\end{equation}
where $J (X)$ is given by the graded expression

$$J (X) \,\, = \,\, {\partial \langle {\cal L} \rangle
\over \partial S_{T} } \, - \, {\partial \over \partial X}
{\partial \langle {\cal L} \rangle \over \partial S_{TX} }
\, + \, {\partial^{2} \over \partial X^{2}}
{\partial \langle {\cal L} \rangle \over \partial S_{TXX}}
\, + \dots \, = \,\, \sum_{s\geq0} (-1)^{s} \,
{\partial^{s} \over \partial X^{s}}
{\partial \langle {\cal L} \rangle \over \partial S_{TsX}} $$

 The Hamiltonian function is given also by the standard 
expression 

$$H [k, J, n] \,\, = \,\, \int_{-\infty}^{+\infty} 
\left( - \, n (X) \, \Sigma_{T} (X) \, + \,
J (X) \, S_{T} (X) \, - \, \langle {\cal L} \rangle
\right) \, d X $$

 Hamiltonian structure (\ref{CanHamStr}) is given in fact by 
the restriction of the Symplectic structure corresponding to
the Gardner - Zakharov - Faddeev bracket to the submanifold
${\cal K}$ and so gives the canonical form of the restriction
of this bracket considered in Theorem 4.2. The functional

$$ I \,\,\, = \,\,\, \int_{-\infty}^{+\infty} J (X) \, d X $$
gives the third annihilator of the restricted
Gardner - Zakharov - Faddeev bracket, so we have here the complete 
set of the canonical variables. Finally, let us say that the
functionals $J (X)$, $H$ should be re-expanded in the graded form
corresponding to the variables $(k, A, n)$ using relation
(\ref{epsdisp}) to come back to our initial gradation rules.
\footnote{Let us note also that relation (\ref{epsdisp})
can not be defined from Lagrangian function and should be 
defined separately from the asymptotic procedure.} 

 Let us say at the end that we believe that the averaging of the 
Symplectic structure is also possible in the case of the Magri
bracket. However, the Symplectic form is much more complicated
in this case and no procedure of such kind of Symplectic forms
is known by now. Let us mention also, that the procedure of the
restriction of Poisson and Symplectic structures must be also
generalized to the so-called weakly non-local structures,
however, we will not discuss here this questions.

\vspace{0.5cm}

 The author is grateful to Prof. B.A. Dubrovin for many fruitful
discussions. 

\vspace{0.5cm}

 The work was supported by the grant
of President of Russian Federation (MD-4903.2008.2) and
RFBR (09-01-92442-KE-a, 09-01-12148-ofi-m).

\end{document}